\documentclass[12pt]{article}

\usepackage[margin=1in,dvips,a4paper,nohead]{geometry}
\usepackage[margin=1em,font=small,justification=centerlast]{caption}
\usepackage{nonfloat}

\usepackage{pstricks}

\psset{unit=1pt, dimen=middle, linewidth=.5pt, arrowsize=3pt 2}

\newdimen \psx
\newdimen \psy

\def\psddots (#1,#2){
  \psx #1\psxunit
  \psy #2\psyunit
  \qdisk(\psx,\psy){.7pt}
  \advance \psx -.2\psxunit
  \advance \psy .2\psyunit
  \qdisk(\psx,\psy){.7pt}
  \advance \psx .4\psxunit
  \advance \psy -.4\psyunit
  \qdisk(\psx,\psy){.7pt}
}

\def\pssddots (#1,#2){
  \psx #1\psxunit
  \psy #2\psyunit
  \qdisk(\psx,\psy){.7pt}
  \advance \psx .2\psxunit
  \advance \psy .2\psyunit
  \qdisk(\psx,\psy){.7pt}
  \advance \psx -.4\psxunit
  \advance \psy -.4\psyunit
  \qdisk(\psx,\psy){.7pt}
}

\usepackage[noBBpl,slantedGreek]{mathpazo}

\usepackage{amsfonts,amsmath,amssymb}

\usepackage{here}
\usepackage{citesort}

\allowdisplaybreaks

\usepackage{array}

\makeatletter
\def\eqnarray{%
   \stepcounter{equation}%
   \def\@currentlabel{\p@equation\theequation}%
   \global\@eqnswtrue
   \m@th
   \global\@eqcnt\z@
   \tabskip\@centering
   \let\\\@eqncr
   $$\everycr{}\halign to\displaywidth\bgroup
       \hskip\@centering$\displaystyle\tabskip\z@skip{##}$\@eqnsel
      &\global\@eqcnt\@ne\hfil$\displaystyle{\hbox{}##\hbox{}}$\hfil
      &\global\@eqcnt\tw@ $\displaystyle{##}$\hfil\tabskip\@centering
      &\global\@eqcnt\thr@@ \hb@xt@\z@\bgroup\hss##\egroup
         \tabskip\z@skip
      \cr
}
\DeclareRobustCommand
  \sddots{\mathinner{\mkern1mu\raise\p@\hbox{.}\mkern2mu
    \raise4\p@\hbox{.}\mkern2mu\raise7\p@
    \vbox{\kern7\p@\hbox{.}}\mkern1mu}}
\makeatother

\def\mbar#1{\kern 0.1em\overline{\kern -0.1em #1 \kern -0.1em}\kern
0.1em}

\def \:{\mskip .5\thinmuskip}

\def \Ad{\mathop{\mathrm{Ad}}}
\def \ad{\mathop{\mathrm{ad}}}

\def \id{\mathop{\mathrm{id}}}

\def\bbR {\mathbb R}
\def\bbC {\mathbb C}
\def\bbZ {\mathbb Z}
\def\calA {\mathcal A}

\def\calF {\mathcal F}
\def\calG {\mathcal G}
\def\calL {\mathcal L}
\def\calM {\mathcal M}

\def\id{\mathop{\mathrm{id}}}

\def\gothg {\mathfrak g}
\def\gothG {\mathfrak G}
\def\gothgl {\mathfrak{gl}}
\def\gothH {\mathfrak H}

\def\gothso {\mathfrak{so}}
\def\gothsl {\mathfrak{sl}}
\def\gothsp {\mathfrak{sp}}

\def\rmD {\mathrm D}
\def\rmd {\mathrm d}
\def\rme {\mathrm e}
\def\rmi {\mathrm i}
\def\rmGL {\mathrm{GL}}
\def\rmO {\mathrm O}
\def\rmSL {\mathrm{SL}}
\def\rmSO {\mathrm{SO}}
\def\rmSp {\mathrm{Sp}}

\title{On $\bbZ$-graded loop Lie algebras, loop groups,
and Toda equations}

\author{Kh. S. Nirov\\
\small \em Institute for Nuclear Research of
the Russian Academy of Sciences\\[-.3em]
\small \em 60th October Anniversary Prospect
7a, 117312 Moscow, Russia\\[.3em]
A. V. Razumov\\
\small \em Institute for High Energy Physics\\[-.3em]
\small \em 142281 Protvino, Moscow Region, Russia}

\date{}

\begin{document}

\maketitle

\begin{abstract}
Toda equations associated with twisted loop groups are considered.
Such equations are specified by $\mathbb Z$-gradations of the
corresponding twisted loop Lie algebras. The classification
of Toda equations related to twisted loop Lie algebras
with integrable $\mathbb Z$-gradations is discussed.
\end{abstract}


\section{Introduction}

A Toda equation is a matrix differential equation of a special form
equivalent to a set of second order nonlinear differential equations.
It is associated to a Lie group and is specified by a $\bbZ$-gradation
of the corresponding Lie algebra \cite{LezSav92,RazSav94,RazSav97a}.
Of certain interest are also higher grading
\cite{Lez85,GerSav95,FerGerSanSav96,BueFerRaz02} and multi-dimensional
\cite{RazSav97c,RazSav97d} generalizations of Toda systems. See also
\cite{FerMirGui95,FerGalHolMir97}, where affine Toda systems were
treated as two-loop WZNW gauged models with potential terms.

When the Lie groups are from the list of the finite dimensional
complex classical Lie groups,
a group-algebraic classification of Toda equations associated with
them was performed in the papers \cite{RazSav97b,RazSavZue99,NirRaz02}
where they were explicitly written in convenient block matrix forms
induced by the $\bbZ$-gradations.

The group-algebraic and the differential-geometry properties of Toda
systems and their physical implications are essentially different
depending on what Lie group, finite or infinite
dimensional, they are associated to. An instructive master example can
be provided by two simplest cases of Toda systems, the Liouville and
the sine-Gordon equations, with their well-studied drastic
differences. In general, for the case of loop Lie groups one deals
with infinite dimensional manifolds \cite{PreSea86}, which may give
rise to additional problems compared to the finite dimensional case.
Say, when one deals with an arbitrary $\bbZ$-gradation of a loop Lie
algebra, one should take care of possible divergences in infinite
series of grading components. A careful examination of loop groups of
complex simple Lie groups and the corresponding $\bbZ$-graded loop Lie
algebras was undertaken in \cite{NirRaz06}. In particular, a useful
notion of integrable
$\bbZ$-gradations was introduced there. On the basis of that
consideration, Toda equations associated with loop groups of complex
classical Lie groups were explicitly described in a
subsequent paper \cite{NirRaz07}. Here we review the papers
\cite{NirRaz06,NirRaz07} skipping the proofs and concentrating mainly
on the investigation logics.

\section{Toda equations associated with loop Lie groups}
\label{s:3}

\subsection{General definition of Toda equation}
\label{s:2.1}

Here $\calM$ denotes either the Euclidean plane $\mathbb R^2$ or the
complex line $\mathbb C$. We denote the standard coordinates on
$\bbR^2$ by $z^-$ and $z^+$. The same notation is used for the
standard complex coordinate on $\mathbb C$ and its complex conjugate,
$z = z^-$ and $\bar{z} = z^+$ respectively. As usual one writes
$\partial_- = \partial/\partial z^-$
and $\partial_+ = \partial/\partial z^+$.

Recall that a Lie algebra $\gothG$ is said to be {\em $\bbZ$-graded\/}
if there is given a representation of $\gothG$ in the form of the
direct sum of subspaces $\gothG_k$ such that
\[
[\gothG_k, \gothG_l] \subset \gothG_{k+l}
\]
for any $k, l \in \bbZ$. This means that any element $\xi$ of $\gothG$
can be uniquely represented as
\[
\xi = \sum_{k \in \bbZ} \xi_k,
\]
where $\xi_k \in \gothG_k$ for each $k \in \bbZ$. In the case when
$\gothG$ is an infinite dimensional Lie algebra we assume that it is
endowed with the structure of a topological vector space and that the
above series converges absolutely.

Let $\calG$ be a Lie group with its Lie algebra $\gothG$ supplied
with a $\bbZ$-gradation. Assume that for some positive
integer $L$ the grading subspaces $\gothG_{-k}$ and $\gothG_{k}$
are trivial whenever $0 < k < L$.
According to the definition of a $\mathbb Z$-gradation,
its zero-grade subspace $\gothG_{0}$ is a subalgebra of
$\gothG$, and one denotes by $\calG_{0}$ the
connected Lie subgroup of $\calG$ corresponding to this
subalgebra.
The {\em Toda equation\/} associated with the Lie group
$\calG$ is a second order nonlinear matrix differential
equation for a smooth mapping $\Xi$ from $\calM$ to
$\calG_{0}$, explicitly of the form\footnote{We assume for simplicity
that $\calG$ is a subgroup of the group formed by invertible elements
of some unital algebra $\calA$. In this case $\gothG$ can be considered
as a subalgebra of the Lie algebra associated with $\calA$. Actually
one can generalize our consideration to the case of an arbitrary Lie
group $\calG$.}
\begin{equation}
\partial_+(\Xi^{-1} \partial_- \Xi) = [\calF_-, \Xi^{-1} \calF_+ \Xi],
\label{e:1}
\end{equation}
see, in particular, the books \cite{LezSav92,RazSav97a}. In this
equation, $\calF_-$ is some fixed mapping from $\calM$ to
$\gothG_{-L}$ and $\calF_+$ is some fixed mapping from $\calM$ to
$\gothG_{+L}$, which satisfy the conditions
\begin{equation}
\partial_+ \calF_- = 0, \qquad \partial_- \calF_+ = 0. \label{e:2}
\end{equation}
When the Lie group $\calG_0$ is abelian, one says that
the corresponding Toda equation is {\em abelian\/}, otherwise one
deals with a {\em non-abelian\/} Toda equation. Remember also that the
Toda equation can be obtained within the differential-geometry
framework, from the zero curvature condition on a flat connection
in the trivial principal fiber bundle $\calM \times \calG \to \calM$
imposing certain grading and gauge conditions~\cite{RazSav97a}.

The authors of the paper \cite{FerMirGui95} consider equations of the
form (\ref{e:1}) for the case when~$L$ does not satisfy the condition
that the subspaces $\gothG_{-k}$ and $\gothG_{+k}$ are trivial
whenever $0 < k < L$.
Actually their assumption does not lead to new Toda equations.
Indeed, consider in this situation the subalgebra of $\gothG'$
of $\gothG$ defined as
\begin{equation}
\gothG' = \bigoplus_{k \in \bbZ} \gothG_{kL} \label{e:r1}
\end{equation}
and the corresponding subgroup $\calG'$ of the group $\calG$. The
expansion (\ref{e:r1}) defines a $\bbZ$-gradation of $\gothG'$ with
the grading subspaces $\gothG'_k = \gothG_{kL}$, $k \in \bbZ$. It is
evident that the Toda equation associated with the Lie group $\calG$
and such choice of the positive integer $L$ can be considered as the
Toda equation associated with the Lie group $\calG'$ and the
choice~$L=1$.

If there is an isomorphism $F$ from a $\bbZ$--graded Lie algebra
$\gothG$ to a $\bbZ$--graded Lie algebra $\gothH$ relating the
corresponding grading subspaces as $\gothH_k = F(\gothG_k)$, one says
that $\mathbb Z$-gradations of $\gothG$ and $\gothH$ are conjugated by
$F$. Actually, having a $\mathbb Z$-graded Lie algebra $\gothG$, one
can induce a $\mathbb Z$-gradation of an $F$-isomorphic Lie algebra
$\gothH$, using $\gothH_k = F(\gothG_k)$ as its grading subspaces.

It is clear that conjugated $\bbZ$-gradations give actually the same
Toda equations. Therefore, to perform a classification of Toda
equations associated with the Lie group $\calG$ one should classify
non-conjugated $\bbZ$-gradations of its Lie algebra $\gothG$. In the
case when $\gothG$ is a complex classical Lie group a convenient
classification of $\bbZ$-gradations was described and the
corresponding Toda equations were presented in the paper
\cite{NirRaz02}, see also \cite{RazSav97b,RazSavZue99}. Here we review
and discuss the corresponding results obtained in the papers
\cite{NirRaz06,NirRaz07} for the case when $\calG$ is a loop group of
a complex classical Lie group.

\subsection{Loop Lie algebras and loop groups}

Let $\gothg$ be a finite dimensional real or complex Lie algebra. The
{\em loop Lie algebra\/} of $\gothg$, denoted $\calL(\gothg)$, is
usually defined as the linear space $C^\infty(S^1, \gothg)$ of smooth
mappings from the circle $S^1$ to $\gothg$ with the Lie algebra
operation defined pointwise. In this paper we define $\calL(\gothg)$
as the linear space $C^\infty_{2 \pi}(\bbR, \gothg)$ of smooth $2
\pi$-periodic mappings of the real line $\bbR$ to $\gothg$ with the
Lie algebra operation again defined pointwise. One can show that these
two Lie algebras are isomorphic. We assume that
$\calL(\gothg)$ is supplied with the structure of a Fr\'echet
space\footnote{A {\em Fr\'echet space\/} is a complete topological
vector space, whose topology is induced by a countable collection of
seminorms.} in such a way that the Lie algebra operation is
continuous, see, for example, \cite{Ham82,Mil84,NirRaz06}. We call
such a Lie algebra a {\em Fr\'echet Lie algebra\/}.

Now, let $G$ be a Lie group with the Lie algebra $\gothg$. The
{\rm loop group\/} of $G$, denoted
$\calL(G)$, is defined alternatively either as the set $C^\infty(S^1,
G)$ of smooth mappings from $S^1$ to $G$ or as the set $C^\infty_{2
\pi}(\bbR, G)$ of smooth $2 \pi$-periodic mappings from $\bbR$ to $G$
with the group law defined in both cases pointwise. In this paper we
adopt the second definition. We assume that $\calL(G)$ is supplied
with the structure of a Fr\'echet manifold modeled on $\calL(\gothg)$
in such a way that it becomes a Lie group, see, for example,
\cite{Ham82,Mil84,NirRaz06}. Here the Lie algebra of the Lie group
$\calL(G)$ is naturally identified with the loop Lie
algebra~$\calL(\gothg)$.

It is not difficult to generalize the above definitions to the case of
twisted loop Lie algebras and loop groups.
Let $A$ be an automorphism of a Lie algebra $\gothg$ satisfying the
relation $A^M = \id\nolimits_\gothg$ for some positive integer $M$.
The {\em twisted loop Lie algebra\/} $\calL_{A,M}(\gothg)$ is a
subalgebra of the loop Lie algebra $\calL(\gothg)$ formed by the
elements $\xi$ which satisfy the equality $\xi(\sigma + 2 \pi / M) =
A(\xi(\sigma))$.
Similarly, given an automorphism $a$ of a Lie group $G$ which
satisfies the relation $a^M = \id_G$, we define the {\em twisted loop
group\/} $\calL_{a,M}(G)$ as the subgroup of the loop group $\calL(G)$
formed by the elements $\rho$ satisfying the equality $\rho(\sigma + 2
\pi/M) = a(\rho(\sigma))$.
The Lie algebra of a twisted loop group $\calL_{a,M}(G)$ is naturally
identified with the twisted loop Lie algebra $\calL_{A,M}(\gothg)$,
where we denote the automorphism of the Lie algebra $\gothg$
corresponding to the automorphism $a$ of the Lie group $G$ by $A$.

It is clear that a loop Lie algebra $\calL(\gothg)$ can be treated as
a twisted loop Lie algebra $\calL_{\id_\gothg, M}(\gothg)$, where $M$
is an arbitrary positive integer. In its turn, a loop group $\calL(G)$
can be treated as a twisted loop group $\calL_{\id_G, M}(G)$, where
$M$ is again an arbitrary positive integer. In the present paper by
loop Lie algebras and loop groups we mean twisted loop Lie algebras
and twisted loop groups.

\subsection{$\bbZ$-gradations of loop Lie algebras and corresponding
Toda equations}

First, let us recall the method used in the paper \cite{NirRaz02} to
describe $\bbZ$-gradations of the Lie algebras of the complex
classical Lie groups. By these groups we mean the Lie groups
$\rmGL_n(\bbC)$, $\rmO_n(\bbC)$ and $\rmSp_n(\bbC)$, whose Lie
algebras are $\gothgl_n(\bbC)$, $\gothso_n(\bbC)$ and
$\gothsp_n(\bbC)$ respectively, see Section \ref{s:3.1}.

We start with a little wider class of Lie algebras. Let $\gothG$ be a
finite dimensional complex simple Lie algebra endowed with a
$\bbZ$-gradation. Define a linear operator $Q$ acting on an element
$\xi \in \gothG$ as
\begin{equation}
Q \: \xi = \sum_{k \in \bbZ} k \: \xi_k, \label{er:2}
\end{equation}
where $\xi_k$, $k \in \bbZ$, are the grading components of $\xi$.  It
is clear that
\[
\gothG_k = \{ \xi \in \gothG \mid Q \: \xi = k \: \xi \}.
\]
Thus, the operator $Q$ completely determines the corresponding
$\bbZ$-gradation. It is called the grading operator generating the
$\bbZ$-gradation under consideration.

One can easily show that $Q$ is a derivation of the Lie algebra
$\gothG$. It is well known that any derivation of a complex simple Lie
algebra is an inner derivation. Therefore, there is a unique element
$q \in \gothG$ such that $Q \xi = [q, \xi]$. Hence, the problem of
classification of $\bbZ$-gradations of $\gothG$ in the case under
consideration is reduced to the problem of classification of the
elements $q \in \gothG$ such that the operator $\ad(q)$ is semisimple
and has only integer eigenvalues. It is known that any
such element belongs to some Cartan subalgebra of $\gothG$. Since all
Cartan subalgebras are conjugated by inner automorphisms of $\gothG$,
to classify the $\bbZ$-gradations of $\gothG$ up to conjugations one
can assume that the element $q$ belongs to some fixed Cartan
subalgebra.

If $\gothG$ is a complex classical Lie algebra it is convenient to
work with the Cartan subalgebra formed by diagonal matrices. After
that the problem of classification of $\bbZ$-gradations of $\gothG$
becomes almost trivial, and its results can be visually represented
with the help of the corresponding block matrix decompositions of the
elements of~$\gothG$ which appear to be very convenient for
description of the Toda systems associated with complex classical Lie
groups \cite{NirRaz02}.

The main lesson here is that it is useful to describe
$\bbZ$-gradations of a Lie algebra~$\gothG$ by their grading operators
being special cases of derivations of $\gothG$. In the case of
infinite dimensional Fr\'echet Lie algebras one should have in mind
that the requirement of continuity is included into the definition of
a derivation.  When this requirement is rejected, it seems impossible
to obtain substantial results on the form of derivations.

Let now $\calG$ be an infinite dimensional Fr\'echet Lie group and
$\gothG$ be its Fr\'echet Lie algebra. For a general $\bbZ$-gradation
of $\gothG$ one cannot use the relation (\ref{er:2}) to define a
linear operator in $\gothG$ because the series in the right hand side
of the relation (\ref{er:2}) may diverge for some $\xi$. We say that
the $\bbZ$-gradation under consideration is generated by grading
operator if this series converges absolutely for every
$\xi \in \gothG$.

In the case when $\gothG = \calL_{A, M}(\gothg)$ the basic example
is the standard $\bbZ$-gradation generated by the grading operator
$Q = - \rmi \rmd/\rmd s$. Here the grading subspaces are
\[
\calL_{A, M}(\gothg)_{k} = \{ \xi \in
\calL_{A,M}(\gothg) \mid \xi = \rme^{\rmi k s} x, \ x \in \gothg, \
A(x) = \rme^{2\pi \rmi k/M} x \}.
\]

Let $\gothG$ be supplied with a $\bbZ$-gradation which is generated by
the grading operator~$Q$. It can be shown that in this case
\[
Q [\xi, \eta] = [Q \xi, \eta] + [\xi, Q \eta]
\]
for any $\xi, \eta \in \gothG$. Hence, if the grading operator is
continuous it is a derivation of~$\gothG$.

Now we restrict our consideration to the case when $\gothG$ is a loop
Lie algebra $\calL_{A,M}(\gothg)$ with $\gothg$ being a finite
dimensional complex simple Lie algebra.
Assume that $\calL_{A,M}(\gothg)$ is endowed with a $\bbZ$-gradation
which is generated by the continuous grading operator~$Q$ which is in
such case a derivation of $\calL_{A,M}(\gothg)$.
The derivations of $\calL_{A,M}(\gothg)$ can be described explicitly
\cite{NirRaz06}, and this allows one to write\footnote{The signs at
the right hand side of the relation (\ref{er:3}) are chosen for
future convenience. The same reason explains explicit appearance
of the imaginary unit.}
\begin{equation}
Q \xi = - \rmi X(\xi) + \rmi [\eta, \xi], \label{er:3}
\end{equation}
where $X$ is a smooth $2\pi/M$-periodic complex vector field on $\bbR$
and $\eta$ is an element of $\calL_{A,M}(\gothg)$. To go further, it
is desirable to show that $X$ is a real vector field. It can be done
if we restrict ourselves to the case of the so-called integrable
$\bbZ$-gradations \cite{NirRaz06}.

We call a $\bbZ$-gradation of a Fr\'echet Lie algebra $\gothG$
integrable if the mapping $\Phi \colon \bbR \times \gothG \to \gothG$
defined by the relation
\[
\Phi(\tau, \xi) = \sum_{k \in \bbZ} \rme^{- \rmi k \tau } \xi_k
\]
is smooth. Here as usual we denote by $\xi_k$ the grading components
of the element~$\xi$. Any such gradation is generated by a continuous
grading operator $Q$ acting on an element $\xi \in \gothG$ as
\begin{equation}
Q \xi = \rmi \left. \frac{\rmd }{\rmd t} \right|_0 \Phi_\xi.
\label{er:4}
\end{equation}
Here $t$ is a standard coordinate on $\bbR$ and the smooth mapping
$\Phi_\xi \colon \bbR \to \gothG$ is defined by the equality
$\Phi_\xi(\tau) = \Phi(\tau, \xi)$. Furthermore, for any fixed $\tau
\in \bbR$ the mapping $\Phi_\tau \colon \xi \in \gothG \mapsto
\Phi(\tau, \xi)$ is an automorphism of $\gothG$, and all such
automorphisms form a one-parameter subgroup of the group of
automorphisms of $\gothG$.

Return again to the case of a loop Lie algebra $\calL_{A, M}(\gothg)$
with $\gothg$ being a finite dimensional complex simple Lie algebra
and assume that it is supplied with an integrable $\bbZ$-gradation.
Using the explicit form of the automorphisms of $\calL_{A, M}(\gothg)$
\cite{NirRaz06} and the equality~(\ref{er:4}) one sees that in this
case the grading operator $Q$ is defined by the relation~(\ref{er:3})
where the vector field $X$ is real. Here one can show that the vector
field $X$ is either a zero vector field or it has no zeros
\cite{NirRaz06}. In the case when $X$ is a zero vector field some
of the grading subspaces are infinite dimensional. Let us restrict
ourselves to the case of $\bbZ$-gradations with finite dimensional
grading subspaces. In this case one can assume that the function
$X(s)$, where $s$ is a standard coordinate on $\bbR$, is positive.
Indeed, in the case when $X(s)$ is negative one can conjugate the
considered $\bbZ$-gradation by the isomorphism sending an element
$\xi \in \calL_{A, M}(\gothg)$ into the element
$\xi' \in \calL_{A^{-1},M}(\gothg)$ such that
$\xi'(\sigma) = \xi(-\sigma)$. It is clear that
this transformation inverses the sign of the vector
field $X$.

Now, using conjugations by isomorphisms, we try to make the grading operator $Q$ given by the relation (\ref{er:3}) as simple as possible. To this end consider a mapping $F$ from $\calL_{A, M}(\gothg)$ into $C^\infty(\bbR, \gothg)$ defined by the equality
\[
F \xi = \rho (f^{-1*} \xi) \rho^{-1},
\]
where $f$ is a diffeomorphism of $\bbR$, and $\rho$ is an element of $C^\infty(\bbR, G)$. The mapping $F$ is injective and can be considered an isomorphism from $\calL_{A, M}(\gothg)$ to $F(\calL_{A, M}(\gothg))$. One can show that
\[
F Q F^{-1} \xi = - \rmi X'(\xi) + \rmi [\rho \eta' \rho^{-1} + X'(\rho)
\rho^{-1}, \xi],
\]
where $X' = f_* X$ and $\eta' = f^{-1*} \eta$.

Choose the diffeomorphism $f$ so that $f_* X = \rmd/\rmd s$. To this end it suffices to define it by the relation
\[
f(\sigma) = \int_{(0, \sigma)} \rmd s/X(s).
\]
It is important here that the vector field $X$ has no zeros. Since the vector field $X$ is $2\pi/M$-periodic, one has
\[
f(\sigma + 2 \pi/M) = f(\sigma) + 2 \pi/M',
\]
where $M'$ is a positive real integer such that
\[
2 \pi / M' = \int_{(0, 2\pi/M)} \rmd s/X(s).
\]
This equality implies, in particular, that
\begin{equation}
\eta'(\sigma + 2 \pi/M') = A(\eta'(\sigma)). \label{er:5}
\end{equation}

Assume now that the mapping $\rho$ is a solution of the equation
\begin{equation}
\rho^{-1} \rmd \rho/ \rmd s = - \eta'. \label{er:6}
\end{equation}
It is well known that this equation always has solutions, all its
solutions are smooth, and if $\rho$ and $\rho'$ are two solutions then
$\rho' = g \rho$ for some $g \in G$. From the relation (\ref{er:5}) it
follows that if $\rho$ is a solution of (\ref{er:6}) then the mapping
$\rho'$ defined by the equality $\rho'(\sigma) = a^{-1} (\rho(\sigma +
2 \pi/M'))$ is also a solution of (\ref{er:6}). Therefore, for some $g
\in G$ one has
\[
\rho(\sigma + 2 \pi/M') = a (g \rho(\sigma)).
\]

It is not difficult to get convinced that with the choice of $f$ and $\rho$ described above the
mapping $F$ maps $\calL_{A, M}(\gothg)$ isomorphically onto the
Fr\'echet Lie algebra $\gothG$ formed by smooth mappings $\xi$ from
$\bbR$ to $\gothg$ satisfying the condition
\[
\xi(\sigma + 2 \pi/ M') = A'(\xi(\sigma)),
\]
where the automorphism $A'$ is defined as $A' = A \circ \Ad(g)$. The
grading operator $F Q F^{-1}$ generating the conjugated gradation of
$\gothG$ is just $- \rmi \rmd/ \rmd s$. It is not difficult to
show~\cite{NirRaz06} that $M'$ is an integer and that $A^{\prime M'} =
\id_\gothg$. It means that $\gothG = \calL_{A', M'}(\gothg)$.

Thus, we see that any integrable $\mathbb Z$-gradation of the loop Lie
algebra $\calL_{A,M}(\gothg)$ with finite dimensional grading
subspaces is conjugated by an isomorphism to the standard gradation of
another twisted loop Lie algebra $\calL_{A',M'}(\gothg)$, where the
automorphisms $A$ and $A'$ differ by an inner automorphism of
$\gothg$. Recall that we consider the case when~$\gothg$ is a
finite dimensional complex simple Lie algebra.

Let $G$ be a finite dimensional complex simple Lie group, $a$ be an
automorphism of $G$ of order $M$, and $A$ be the corresponding
automorphism of the Lie algebra $\gothg$ of $G$. To classify Toda
equations associated with $\calL_{a, M}(G)$ one should classify, up to
conjugation by isomorphisms, $\mathbb Z$-gradations of $\calL_{A,
M}(\gothg)$. As was actually shown above, if we restrict ourselves to
integrable $\bbZ_M$-gradations with finite dimensional grading
subspaces this task is equivalent to classification of the loop groups
$\calL_{a, M}(G)$ themselves, or, equivalently, to classification of
finite order automorphisms of $G$. It is not difficult to get
convinced \cite{NirRaz07} that it suffices to perform the latter
classification also up to conjugation by isomorphisms. As a matter of
fact, we will classify finite order automorphisms of the Lie algebra
$\gothg$ which can be lifted to automorphisms of the Lie group $G$.

It is very useful to realize that every automorphism $A$ of $\gothg$
satisfying the relation
$A^M = \mathrm{id}_\gothg$ induces a $\mathbb Z_M$-gradation of
$\gothg$ with the grading subspaces\footnote{We denote by $[k]_M$ the
element of the ring $\mathbb Z_M$ corresponding to the integer $k$.}
\[
\gothg_{[k]_M} = \{x \in \gothg \mid A(x) = \rme^{2 \pi \rmi k / M}
x\}, \qquad k = 1, \ldots, M-1.
\]
Vice versa, any $\mathbb Z_M$-gradation of $\gothg$ defines in an
evident way an automorphism $A$ of $\gothg$ satisfying the relation
$A^M = \mathrm{id}_\gothg$. A $\mathbb Z_M$-gradation of $\gothg$ is
called an inner or outer type gradation, if the associated
automorphism $A$ of $\gothg$ is of inner or outer type respectively.

The grading subspaces of the standard $\mathbb Z$-gradation of the
twisted loop Lie algebra $\calL_{A,M}(\gothg)$ can be described in
terms of the corresponding $\mathbb Z_M$-gradation
of $\gothg$ as
\[
\calL_{A,M}(\gothg)_k = \{ \xi \in \calL_{A,M}(\gothg)  \mid \xi =
\rme^{\rmi k s} x, \ x \in \gothg_{[k]_M} \}.
\]

For the standard gradation the positive integer $L$ entering the
definition of a Toda equation satisfies the inequality $L \le M$, and
the equality $L = M$ takes place
if and only if $A = \mathrm{id}_\gothg$, with the positive integer $M$
being arbitrary. In this case the nontrivial grading subspaces are
$\calL_{{\mathrm{id}_\gothg},M}(\gothg)_{kM}$
for $k \in \mathbb Z$, and one has
\[
\calL_{\mathrm{id}_\gothg,M}(\gothg)_{kM}
= \{ \xi \in \calL_{\mathrm{id}_\gothg,M}(\gothg)
 \mid \xi = \rme^{\rmi k M s} x, \ x \in \gothg \}.
\]

It is also clear that for the standard $\bbZ$-gradation the subalgebra
$\calL_{A, M}(\gothg)_0$ is isomorphic to $\gothg_{[0]_M}$, and the
Lie group $\calL_{a, M}(G)_0$ is isomorphic to the connected Lie
subgroup $G_0$ of $G$ corresponding to the Lie algebra
$\gothg_{[0]_M}$. Hence, the mapping $\Xi$ is actually a mapping
from $\calM$ to $G_0$, for consistency with the notation used earlier
we will denote it by $\gamma$. The mappings $\calF_-$ and $\calF_+$
are given by the relation
\[
\calF_-(p) = \rme^{- \rmi L s} c_-(p), \qquad \calF_+(p) = \rme^{\rmi L s}
c_+(p),
\qquad p \in \calM,
\]
where $c_-$ and $c_+$ are
mappings from $\calM$ to $\gothg_{-[L]_M}$ and $\gothg_{+[L]_M}$
respectively. Thus, the Toda equation (\ref{e:1}) can be written as
\begin{equation}
\partial_+(\gamma^{-1} \partial_- \gamma) = [c_-, \gamma^{-1} c_+
\gamma], \label{e:5}
\end{equation}
where $\gamma$ is a smooth mapping from $\calM$ to $G_0$, and the
mappings $c_-$ and $c_+$ are fixed smooth
mappings from $\calM$ to $\gothg_{-[L]_M}$ and $\gothg_{+[L]_M}$
respectively. The conditions (\ref{e:2}) imply that
\begin{equation}
\partial_+ c_- = 0, \qquad \partial_- c_+ = 0. \label{e:6}
\end{equation}

Summarizing one can say that a Toda equation associated with a loop
group of a simple complex Lie group whose Lie algebra is endowed with
an integrable $\bbZ$-gradation with finite dimensional grading
subspaces is equivalent to the equation of the form~(\ref{e:5}).

It is reasonable to single out the simplest case, with $A$ being
$\id_\gothg$ and $M$ an arbitrary positive number. In this case $L =
M$. The mapping $\gamma$ is a mapping from $\calM$ to the whole group
$G$, and $c_+$ and $c_-$ are mappings from $\calM$ to $\gothg$.
Denoting in this particular case $\gamma$ by $\Gamma$, $c_+$ by $C_+$,
and $c_-$ by $C_-$, one writes the Toda equation
(\ref{e:5}) as
\begin{equation}
\partial_+(\Gamma^{-1} \partial_- \Gamma)
= [C_-, \Gamma^{-1} C_+ \Gamma],
\label{e:5a}
\end{equation}
and the conditions (\ref{e:6}) as
\begin{equation}
\partial_+ C_- = 0, \qquad \partial_- C_+ = 0.
\label{e:6a}
\end{equation}
Note that this example indicates a principal difference between Toda
systems associated with finite dimensional and loop Lie groups.

One can consider equations of the type (\ref{e:5}) in a more general
setting. Namely, let $G$ be an arbitrary finite dimensional Lie group
and $a$ be an arbitrary finite order
automorphism of $G$. The corresponding automorphism $A$ of the Lie
algebra $\gothg$ of the Lie group $G$ generates a $\bbZ_M$-gradation
of $\gothg$. Assume that for some positive integer $L \le M$ the
grading subspaces $\gothg_{+[k]_M}$ and $\gothg_{-[k]_M}$ for $0 < k <
L$ are trivial. Choose some fixed mappings $c_+$ and $c_-$ from
$\calM$ to $\gothg_{+[L]_M}$ and $\gothg_{-[L]_M}$, respectively,
satisfying the relations (\ref{e:6}). Now the equation (\ref{e:5}) is
equivalent to a Toda equation associated with the loop group
$\calL_{a, M}(G)$ whose Lie algebra $\calL_{A, M}(\gothg)$ is endowed
with the standard $\bbZ$-gradation.

Note that the authors of the paper \cite{FerGalHolMir97} also suggest
the equation (\ref{e:5}) as a convenient form of a Toda equation
associated with a loop group. They do not assume that the grading
subspaces $\gothg_{+[k]_M}$ and $\gothg_{-[k]_M}$ for $0 < k < L$
are trivial. Actually their assumption does not lead to new Toda
equations.

Indeed, suppose that we do not assume that the grading subspaces
$\gothg_{+[k]_M}$ and $\gothg_{-[k]_M}$ for $0 < k < L$ are
trivial. As in Section \ref{s:2.1} consider a subalgebra
\[
\gothG' = \bigoplus_{k \in Z} \calL_{A, M}(\gothg)_{kL}
\]
of the loop Lie algebra $\calL_{A, M}(\gothg)$ as a $\bbZ$-graded Lie
algebra with the grading subspaces $\gothG'_k = \calL_{A,
M}(\gothg)_{kL}$. The Toda equation associated with the Lie group
$\calL_{a, M}(G)$ and the above choice of the positive integer $L$ can
be considered as the Toda equation associated with the Lie group
$\calG'$ corresponding to the Lie algebra $\gothG'$ and the
choice~$L = 1$. Let us show that the Lie algebra $\gothG'$ is
isomorphic to some loop Lie algebra.

Let $M'$ be the minimal positive integer such that $M' [L]_M = [0]_M$.
In other words, $M'$ is the order of the element $[L]_M$ considered as
an element of the additive group of the ring $\bbZ_M$.
Consider the subalgebra $\gothg'$ of the Lie algebra $\gothg$ defined
as
\[
\gothg' = \bigoplus_{k=0}^{M'-1} \gothg_{k[L]_M}.
\]
The Lie algebra $\gothg'$ can be treated as a $\bbZ_{M'}$-graded Lie
algebra with the grading subspaces $\gothg'_{[k]_{M'}} =
\gothg_{k[L]_M}$. Denote the corresponding automorphism of $\gothg'$
by $A'$.

Every element $\xi \in \gothG'$ can be represented as the absolutely
convergent sum
\[
\xi = \sum_{k \in \bbZ} \rme^{\rmi k L s} x_k,
\]
where for each $k \in \bbZ$ one has $x_k \in \gothg_{k[L]_M}$. It is
clear that the series
\[
\xi' = \sum_{k \in \bbZ} \rme^{\rmi k s} x_k
\]
is absolutely convergent, and $\xi'$ can be considered as an element
of $\calL_{A', M'}(\gothg)$. It can be easily verified that the
mapping sending $\xi$ to $\xi'$ is an isomorphism from $\gothG'$ to
$\calL_{A', M'}(\gothg)$.

Thus, if we do not not assume that the grading subspaces
$\gothg_{+[k]_M}$ and $\gothg_{-[k]_M}$ for $0 < k < L$ are trivial,
then the arising Toda equation associated with the loop group
$\calL_{A, M}(G)$ can be considered as the Toda equation associated
with some other loop group $\calL_{A', M'}(G')$ for the case when $L =
1$. Here $G'$ is the Lie group corresponding to the subalgebra
$\gothg'$ of the Lie algebra $\gothg$.

In what follows, explicit forms which the Toda equation (\ref{e:5})
takes for complex classical Lie groups $G$ are specified. It was
pointed out that this specification should use the classification,
up to conjugations, of the finite order automorphisms of the Lie
algebras under consideration. Instead of using root techniques
as, for example, in \cite{OniVin90,Kac94,GorOniVin94}, here the
classification in terms of convenient block matrix
representations is implemented.

\section{$\bbZ_M$-gradations of complex classical Lie algebras}
\label{s:4}

\subsection{Complex classical Lie groups and Lie algebras}
\label{s:3.1}

Here we define the complex classical Lie groups and discuss their
basic properties. More information on these groups can be found, for
example, in the works \cite{OniVin90,GorOniVin94,NirRaz07}.

Let us first explain the notation used. We denote by $I_n$ the unit
diagonal $n \times n$ matrix and by $J_n$ the symmetric skew diagonal
$n \times n$ matrix. For an even $n$ we also define the skew
symmetric skew diagonal $n \times n$ matrix
\[
K_n = \left( \begin{array}{cc}
0 & J_{n/2} \\
- J_{n/2} & 0
\end{array} \right).
\]
When it does not lead to a misunderstanding, we write instead of
$I_n$, $J_n$, and $K_n$ just $I$, $J$, and $K$ respectively.

Besides, the following convention is being used.
If $m$ and $B$ be $n \times n$ matrices, one denotes
${}^{B\!} m = B^{-1} \: {}^{t\!} m \: B$,
where ${}^{t\!} m$ is the transpose of the matrix $m$. Note that
${}^{J\!} m$ is
actually the transpose of $m$ with respect to the skew diagonal.

The complex general linear group $\rmGL_n(\bbC)$ is formed by all
nonsingular complex \hbox{$n \times n$} matrices with matrix
multiplication as the group law. The Lie algebra $\gothgl_n(\bbC)$ of
the Lie group $\rmGL_n(\bbC)$ is formed by all complex $n \times n$
matrices with matrix commutator as the Lie algebra law. The Lie group
$\rmGL_n(\bbC)$ and the Lie algebra $\gothgl_n(\bbC)$ are not simple.
The subgroup $\rmSL_n(\bbC)$ of $\rmGL_n(\bbC)$ formed by the
matrices with unit determinant is called the complex special
linear group. The Lie group
$\rmSL_n(\bbC)$ is connected and simple. Its Lie algebra
$\gothsl_n(\bbC)$ is also simple.

Let $B$ be a complex nonsingular $n \times n$
matrix. The elements $g$ of $\rmGL_n(\bbC)$ singled out by the
condition ${}^{B\!} g = g^{-1}$ form a Lie subgroup of
$\rmGL_n(\bbC)$ which one denotes by $\rmGL_n^B(\bbC)$. The Lie
algebra $\gothgl_n^B(\bbC)$ of $\rmGL_n^B(\bbC)$ is a subalgebra
of the Lie algebra $\gothgl_n(\bbC)$ formed by the complex
$n \times n$ matrices $x$ satisfying the condition
${}^{B\!} x = -x$.

For any symmetric nonsingular $n \times n$ matrix $B$ the
Lie group $\rmGL_n^B(\bbC)$ is isomorphic to the Lie group
$\rmGL_n^J(\bbC)$. This group is called the complex orthogonal
group and is denoted $\rmO_n(\bbC)$. For an element
$g \in \rmO_n(\bbC)$ from the equality ${}^{J\!} g = g^{-1}$
one obtains that $\det g$ is equal either to $1$ or to $-1$.
The elements of $\rmO_n(\bbC)$ with unit determinant form a
connected Lie subgroup of $\rmO_n(\bbC)$ called the complex
special orthogonal group and is denoted $\rmSO_n(\bbC)$. This
subgroup is the connected component of the identity of
$\rmO_n(\bbC)$. The Lie algebra of $\rmSO_n(\bbC)$ is denoted
$\gothso_n(\bbC)$. It is clear that the Lie algebra of
$\rmO_n(\bbC)$ coincides with the Lie algebra of $\rmSO_n(\bbC)$.
The Lie group $\rmSO_n(\bbC)$ and the Lie algebra $\gothso_n(\bbC)$
are simple. The special orthogonal Lie group and its Lie algebra
possess both inner and outer automorphisms, the latter exist only when
$n$ is even.

For an even $n$ take a skew symmetric nonsingular $n \times n$ matrix
$B$. In this case the Lie group $\rmGL_n^B(\bbC)$ is isomorphic to
the Lie group $\rmGL_n^K(\bbC)$. This group is called the complex
symplectic group and is denoted $\rmSp_n(\bbC)$. The Lie group
$\rmSp_n(\bbC)$ is connected and simple. The corresponding Lie
algebra $\gothsp_n(\mathbb C)$ is also simple. Remember that
the symplectic Lie group and its Lie algebra have only inner
automorphisms.

\subsection{$\bbZ_M$-gradations of complex general linear Lie algebras
of inner type}
\label{s:4.1}

In this section we consider inner type $\bbZ_M$-gradations of complex general linear Lie algebras. We call such gradations gradations of $\gothgl_n(\bbC)$ of type~I. There are two more types of $\bbZ_M$-gradations generated by outer automorphisms of complex general linear Lie algebras. They will be considered in Section~\ref{s:5.1}.

\subsubsection*{Type I}

Let $a$ be an inner automorphism of the Lie group $\rmGL_n(\bbC)$
satisfying the relation $a^M = \id_{\rmGL_n(\bbC)}$. Denote the
corresponding inner automorphism of the Lie algebra $\gothgl_n(\bbC)$
by $A$. This automorphism satisfies the relation
$A^M = \id_{\gothgl_n(\bbC)}$. In other words, $A$ is a finite
order automorphism of $\gothgl_n(\bbC)$. Since one is interested
in the automorphisms of $\gothgl_n(\bbC)$ up to conjugations, it
can be assumed that the automorphism $A$ under consideration is given
by the relation
\begin{equation}
A(x) = h \: x \: h^{-1}, \label{er:7}
\end{equation}
where $h$ is an element of the subgroup $\rmD_n(\bbC)$ of
$\rmGL_n(\bbC)$ formed by all complex nonsingular diagonal
matrices, see, for example, \cite{NirRaz07}. It is clear that
multiplying $h$ by an arbitrary
nonzero complex number one obtains an element of $\rmD_n(\bbC)$
which generates the same automorphism of $\gothgl_n(\mathbb C)$
as the initial element.

The equality $A^M = \id_{\gothgl_n(\bbC)}$ gives
$h^M \: x \: h^{-M} = x$ for any $x \in \gothgl_n(\mathbb C)$.
Therefore, $h^M = \nu I$, where $\nu$ is a nonzero complex number.
Multiplying $h$ by an appropriate complex number we make it satisfy
the relation $h^M = I$. This means that the diagonal matrix elements
of $h$ have the form $\rme^{2 \pi \rmi m/M}$, where $m$ is an integer.
We will assume that $0 < m \le M$. Using inner automorphisms of
$\gothgl_n(\mathbb C)$ which permute the rows and columns of the
matrix $h$ synchronously, we collect coinciding diagonal matrix
elements together, and come to the following block diagonal form of
the element $h$:
\begin{equation}
\psset{xunit=2.3em, yunit=1.4em}
h = \left( \raise -1.6\psyunit \hbox{\begin{pspicture}(.4,.7)(4.7,4.5)
\rput(1,4){$\mu_1 I_{n_1}$} \rput(2,3){$\mu_2 I_{n_2}$} \psddots(3,2)
\rput(4,1){$\mu_p I_{n_p}$}
\end{pspicture}} \right). \label{e:9}
\end{equation}
Here $\mu_\alpha= \rme^{2 \pi \rmi m_\alpha / M}$, the positive
integers $m_\alpha$ form a decreasing sequence, $M \ge m_1 > m_2 >
\ldots > m_p> 0$, and the positive integers $n_\alpha$ satisfy the
equality $\sum^p_\alpha n_\alpha = n$. It is assumed that the integer
$p$ is greater than $1$. The case $p = 1$ corresponds to $A =
\mathrm{id}_{\gothgl_n(\bbC)}$, and one reveals the equation
(\ref{e:5a}), where $\Gamma$ is a mapping from $\calM$ to
$\mathrm{GL}_{n}(\mathbb C)$, and $C_-$ and $C_+$ are mappings from
$\calM$ to $\gothgl_n(\bbC)$ satisfying the conditions (\ref{e:6a}).

Now consider the corresponding $\bbZ$-gradation. Represent
the general element $x$ of $\gothgl_n(\mathbb C)$ in the
block matrix form suggested by the structure of $h$,
\begin{equation}
x = \left( \begin{array}{cccc}
x_{11} & x_{12} & \cdots & x_{1p} \\[.5em]
x_{21} & x_{22} & \cdots & x_{2p} \\[.5em]
\vdots & \vdots & \ddots & \vdots \\[.5em]
x_{p1} & x_{p2} & \cdots & x_{pp}
\end{array} \right), \label{e:10}
\end{equation}
where $x_{\alpha \beta}$, $\alpha, \beta = 1, \ldots, p$, is an
$n_\alpha \times n_\beta$ matrix.
One easily finds
\[
(h \: x \: h^{-1})_{\alpha \beta}^{}
= \rme^{2 \pi \rmi (m_\alpha - m_\beta) / M} x_{\alpha \beta}.
\]
Hence, if for fixed $\alpha$ and $\beta$ only the block
$x_{\alpha \beta}$ of the element $x$ is different from zero,
then $x$ belongs to the grading subspace $[m_\alpha - m_\beta]_M$.
It is convenient to introduce integers
$k_\alpha$, $\alpha = 1, \ldots, p-1$, defined as
$k_\alpha = m_\alpha - m_{\alpha + 1}$. By definition,
for each $\alpha$ the integer $k_\alpha$ is positive
and $\sum_{\alpha=1}^{p-1} k_\alpha = m_1 - m_p < M$.
It is clear that for $\alpha < \beta$ one has
$[m_\alpha - m_\beta]_M
= [\sum_{\gamma=\alpha}^{\beta-1} k_\gamma]_M$,
and for $\alpha > \beta$ one has
$[m_\alpha - m_\beta]_M =
- [\sum_{\gamma=\beta}^{\alpha-1} k_\gamma]_M =
[M - \sum_{\gamma=\beta}^{\alpha-1} k_\gamma]_M$.
These relations allow one to describe the grading structure
of the $\bbZ_M$-gradation generated by the automorphism $A$ by
Figure \ref{f:1}.
\begin{figure}[htb]
\[
\left( \begin{array}{c|c|c|c|c}
[0]_M & [k_1]_M & [k_1 + k_2]_M & \cdots & \displaystyle
[\sum_{\alpha=1}^{p-1} k_\alpha]_M \\ \hline
-[k_1]_M & [0]_M & [k_2]_M & \cdots & \displaystyle
[\sum_{\alpha=2}^{p-1} k_\alpha]_M \\ \hline
-[k_1 + k_2]_M & - [k_2]_M & [0]_M & \cdots & \displaystyle
[\sum_{\alpha = 3}^{p-1} k_\alpha]_M \\ \hline
\vdots & \vdots & \vdots & \ddots & \vdots \\ \hline
\displaystyle -[\sum_{\alpha = 1}^{p-1} k_\alpha]_M &
\displaystyle
- [\sum_{\alpha=2}^{p-1} k_\alpha]_M &
\displaystyle - [\sum_{\alpha=3}^{p-1}
k_\alpha]_M & \cdots & [0]_M
\end{array} \right)
\]
\caption{The canonical structure of a $\bbZ_M$-gradation}
\label{f:1}
\end{figure}
Here the elements of the ring $\bbZ_M$ are the grading indices of
the corresponding blocks in the block matrix representation
(\ref{e:10}) of a general element of $\mathfrak{gl}_n(\mathbb C)$.
Note, in particular, that the subalgebra $\gothg_{[0]_M}$ is formed
by all block diagonal matrices and is isomorphic to the Lie algebra
$\gothgl_{n_1}(\mathbb C) \times \cdots
\times \gothgl_{n_p}(\mathbb C)$.
The group $G_0$ is also formed by block diagonal matrices, and as
such, it is isomorphic to
$\mathrm{GL}_{n_1}(\mathbb C) \times \cdots \times
\mathrm{GL}_{n_p}(\mathbb C)$. This convenient structure of
$\gothg_{[0]_M}$ and $G_0$ is due to the chosen ordering of the
diagonal matrix elements of $h$. Here one can say that the numbers
$\mu_\alpha$ are ordered clock wise as points on the unit circle in
the complex plane.

It is clear that up to conjugations a $\bbZ_M$-gradation of
$\gothgl_n(\bbC)$ of inner type can be specified by a choice of $p \le
n$ positive integers $n_\alpha$, satisfying the equality
$\sum_{\alpha=1}^p n_\alpha = n$, and $p-1$ positive integers
$k_\alpha$, satisfying the inequality $\sum_{\alpha = 1}^{p-1}
k_\alpha < M$. The corresponding automorphism of $\gothgl_n(\bbC)$ is
defined by the relation (\ref{er:7}), where $h$ is given by the
equality (\ref{e:9}). Here $\mu_\alpha = \rme^{2 \pi \rmi m_\alpha /
M}$ with
\begin{equation}
m_\alpha = \sum_{\beta = \alpha}^{p-1} k_\beta + m_p, \qquad \alpha =
1, \ldots, p-1, \label{er:8}
\end{equation}
while $m_p$ is an arbitrary positive integer such that the inequality
$\sum_{\alpha = 1}^{p-1} k_\alpha < M$ is valid. Different choices of
$m_p$ give the same automorphism of $\gothgl_n(\bbC)$.

\subsection{$\bbZ_M$-gradations of complex orthogonal and symplectic
Lie algebras} \label{s:4.3}

Up to conjugations, all inner and outer type $\bbZ_M$-gradations of
complex orthogonal and symplectic Lie algebras are generated by
automorphisms given by the relation~(\ref{er:7}),\footnote{Strictly
speaking, for the Lie algebra $\gothso_8(\bbC)$ there are outer
automorphisms which are not described by the relation (\ref{er:7}).
However, these automorphisms cannot be lifted up to automorphisms of
the Lie group $\rmSO_8(\bbC)$ and are not relevant for our purposes.}
where $h$ belongs either to the Lie group $\rmO_n(\bbC) \cap
\rmD_n(\bbC)$ or to the Lie group $\rmSp_n(\bbC) \cap \rmD_n(\bbC)$.
Actually $\rmO_n(\bbC) \cap \rmD_n(\bbC) = \rmSp_n(\bbC) \cap
\rmD_n(\bbC)$ and it is convenient to assume that $h \in \rmO_n(\bbC)
\cap \rmD_n(\bbC)$. For any element $x \in \gothgl^B_n(\bbC)$ one has
$h^M \: x \: h^{-M} = x$, therefore, $h^M = \nu I$ for some complex
number $\nu$. This equality implies that $({}^{B\!} h)^M = \nu I$,
therefore, $({}^{B\!} h)^M \: h^M = \nu^2 I$. From the other hand,
using the equality ${}^{B\!} h \: h = I$, one obtains
$({}^{B\!} h)^M \: h^M = ({}^{B\!} h \: h)^M = I$. Thus, one sees that
$\nu^2 = 1$. In other words, one has that $\nu$ is equal either to $1$
or to $-1$. In both cases the diagonal matrix elements of $h$ are of
modulus one.

To come to the canonical structure of a $\bbZ_M$-gradation one has to
bring $h$ to the form (\ref{e:9}) where the numbers $\mu_\alpha$ are
ordered clock wise as points on the unit circle in the complex plane.
It appears that in some cases this cannot be done by automorphisms of
the Lie algebra under consideration. Actually in these cases some
diagonal matrix elements are equal to $1$ and some of them are equal
to $-1$ and one cannot collect such elements together keeping $h$
in $\gothso_n(\bbC)$ or $\gothsp_n(\bbC)$. It may seem that similar
obstructions for the required ordering arise also in the case when
some diagonal matrix elements are equal to $1$ even if there are no
matrix elements equal to $-1$. However, in this case one can multiply
$h$ by $-1$ thus overcoming the problem. As a result we obtain two types of $\bbZ_M$-gradations of complex orthogonal and symplectic Lie algebras.

\subsubsection*{Type I}

We start with the case when it is possible to perform the desired ordering of
the numbers $\mu_\alpha$ staying within $\gothso_n(\bbC)$ or
$\gothsp_n(\bbC)$.
As for the case of $\gothgl_n(\bbC)$ a $\bbZ_M$-gradation of
$\gothso_n(\bbC)$ or $\gothsp_n(\bbC)$ can be specified by a choice of
$p \le n$ positive integers $n_\alpha$, satisfying the equality
$\sum_{\alpha=1}^p n_\alpha = n$, and \hbox{$p-1$} positive integers
$k_\alpha$, satisfying the inequality $\sum_{\alpha = 1}^{p-1}
k_\alpha < M$. But obviously these integers are no more as arbitrary
as in the general linear case. They are subject to additional
constraints coming from the corresponding
Lie group and algebra defining conditions. Explicitly, the numbers
$n_\alpha$ satisfy the equalities
\[
n_{p - \alpha + 1} = n_\alpha, \qquad \alpha = 1, \ldots, p,
\]
and for the numbers $k_\alpha$ one has
\[
k_{p - \alpha} = k_\alpha, \qquad \alpha = 1, \ldots, p - 1.
\]

Let $M - \sum_{\alpha=1}^{p-1} k_\alpha$ be an even positive integer.
In this case the automorphism of $\gothso_n(\bbC)$ or
$\gothsp_n(\bbC)$ generating the $\bbZ_M$-gradation under
consideration is defined by the relation (\ref{er:7}) with $h$ given
by the equality (\ref{e:9}) where $\mu_\alpha = \rme^{2 \pi \rmi
m_\alpha /M}$. The integer $m_p$ is now fixed by the equality
\[
m_p = \frac{1}{2}\left( M - \sum_{\alpha=1}^{p-1} k_\alpha \right),
\]
and the integers $m_\alpha$, $\alpha = 1, \ldots, p-1$, are given by
the relation (\ref{er:8}).

Let now $M - \sum_{\alpha=1}^{p-1} k_\alpha + 1$ be an even positive
integer. In this case the automorphism of $\gothso_n(\bbC)$ or
$\gothsp_n(\bbC)$ generating the $\bbZ_M$-gradation under
consideration is defined by the relation (\ref{er:7}) with $h$ given
by the equality (\ref{e:9}) where $\mu_\alpha = \rme^{2 \pi \rmi
(m_\alpha + 1/2) /M}$. The integer $m_p$ is fixed by the equality
\[
m_p = \frac{1}{2}\left( M - \sum_{\alpha=1}^{p-1} k_\alpha + 1
\right),
\]
and the integers $m_\alpha$, $\alpha = 1, \ldots, p-1$, are again
given by the relation (\ref{er:8}).

In both cases the structure of the $\bbZ_M$-gradation under
consideration is depicted by Figure~\ref{f:1}. However, now one is
faced with the necessity to impose the appropriate restrictions on
the blocks of the decomposition (\ref{e:10}). This implies in
particular that the Lie algebra $\gothg_{[0]_M}$ is formed by block
diagonal matrices. This Lie algebra is isomorphic to
$\gothgl_{n_1}(\bbC) \times \cdots \times \gothgl_{n_s}(\bbC)$ for
an even $p = 2s$, and it is isomorphic to
$\gothgl_{n_1}(\bbC) \times \cdots \times
\gothgl_{n_{s-1}}(\bbC) \times \gothgl^{B_s}_{n_s}(\bbC)$ for an odd
$p = 2s - 1$. Here either $B_s = J_{n_s}$ or $B_s = K_{n_s}$ depending
on what kind of Lie algebras is considered. It is clear that in the
symplectic case for an odd $p$ the integer $n_s$ should be even. The
Lie group $G_0$ is isomorphic either to $\rmGL_{n_1}(\bbC) \times
\cdots \times \rmGL_{n_s}(\bbC)$ for an even $p = 2s$, or to
$\rmGL_{n_1}(\bbC) \times \cdots \times \rmGL_{n_{s-1}}(\bbC) \times
\rmGL^{B_s}_{n_s}(\bbC)$ for an odd $p = 2s - 1$.

\subsubsection*{Type II}

This type corresponds to the case when some diagonal matrix elements
of the element $h$ generating the $\bbZ_M$-gradation under
consideration are equal to $1$ and some of them are equal to $-1$.
Note that this is possible only if $M$ even and also $p$ is even.

Permuting the rows and columns of the matrices representing the
elements of the Lie algebra under consideration in an appropriate way
we move the diagonal matrix elements of $h$ equal to $1$ to the
beginning of the diagonal. This transformation can be performed in
such a way that the Lie algebra is mapped isomorphically onto the Lie
algebra $\gothgl^B_n(\bbC)$ with
\[
B = \left( \begin{array}{cc}
J_{n_1} & 0 \\
0 & J_{n-n_1}
\end{array} \right)
\]
in the orthogonal case, and
\[
B = \left( \begin{array}{cc}
K_{n_1} & 0 \\
0 & K_{n-n_1}
\end{array} \right)
\]
in the symplectic case. Here $n_1$ is the number of diagonal
matrix elements of $h$ equal to~$1$. Then by an automorphism of
$\gothgl^B_n(\bbC)$ we order the remaining diagonal matrix elements
of $h$ clock wise on the unit circle in the complex plane. As the
result of our transformation we come to $\bbZ_M$-gradations which
are specified by a set of $p \le n$ positive integers $n_\alpha$
such that $\sum_{\alpha=1}^p n_\alpha = n$ and
\[
n_{p - \alpha + 2} = n_\alpha, \qquad \alpha = 2, \ldots, p,
\]
and by a set of $p-1$ positive integers $k_\alpha$ such that
\[
\sum_{\alpha = 1}^{p - 1} k_\alpha + k_1 = M
\]
and
\[
k_{p - \alpha + 1} = k_\alpha, \qquad \alpha = 2, \ldots, p - 1.
\]
The automorphism generating the $\bbZ_M$-gradation under consideration
is defined by the relation (\ref{er:7}) with $h$ given by the equality
(\ref{e:9}) where $\mu_\alpha = \rme^{2 \pi \rmi m_\alpha /M}$. Here
$m_p = k_1$ and the integers $m_\alpha$, $\alpha = 1, \ldots, p-1$,
are given by the relation (\ref{er:8}).

The structure of the $\bbZ_M$-gradation under consideration is
depicted by Figure~\ref{f:1}. For $p = 2 s - 2$ the Lie algebra
$\gothg_{[0]_M}$ is isomorphic to $\gothgl_{n_1}^{B_1} \times
\gothgl_{n_2}(\bbC) \times \cdots \times \gothgl_{n_{s-1}}(\bbC)
\times \gothgl^{B_s}_{n_s}(\bbC)$. Here either $B_1 = J_{n_1}$, $B_s =
J_{n_s}$ or $B_1 = K_{n_1}$, $B_s = K_{n_s}$ depending on what kind of
Lie algebras is considered. It is clear that in the symplectic case
the integers $n_1$ and $n_s$ should be even. The Lie group $G_0$
is isomorphic to $\rmGL^{B_1}_{n_1}(\bbC) \times \rmGL_{n_2}(\bbC)
\times \cdots \times \rmGL_{n_{s-1}}(\bbC) \times
\rmGL^{B_s}_{n_s}(\bbC)$.

\subsection{$\bbZ_M$-gradations of complex general linear Lie algebras
of outer type}
\label{s:5.1}

The consideration of this case is based on the observation that an
arbitrary outer automorphism $A$ of $\gothgl_n(\bbC)$ is conjugated to
the automorphism defined by the relation
\[
A(x) = - h \: {}^{J\!} x \: h^{-1},
\]
where $h$ is an element of $\rmGL_n(\bbC) \cap \rmD_n(\bbC)$ such that
${}^{J\!} h \: h = I$. One can show that without loose of generality
one can assume that $\det h = 1$. As in the orthogonal and symplectic
cases $h^M = \nu I$, where $\nu$ is equal either to $1$ or to $-1$.
Furthermore, one can get convinced that outer type $\bbZ_M$-gradations
of $\gothgl_n(\bbC)$ exist only for even $M$. We denote $M/2$ by $N$.

In the simplest case $h = I$, and there are only two nontrivial
grading subspaces $\gothg_{[0]_{2N}}$ and $\gothg_{[N]_{2N}}$ whose
elements are singled out by the conditions
${}^{J\!} x = -x$ and ${}^{J\!} x = x$, respectively. In this case,
the Lie group $G_0$ coincides with $\rmSO_n(\bbC)$ and one comes to
the Toda equation (\ref{e:5a}) with $\Gamma$ being a mapping from
$\cal M$ to $\rmSO_n(\bbC)$, and $C_+$ and $C_-$ being mappings from
$\calM$ to the space of $n \times n$ complex matrices $x$ satisfying
the equality ${}^{J\!} x = x$.

To analyze a general case one uses the following simple observations.
Let $B$ be an arbitrary nonsingular matrix. For any $h \in
\rmGL_n(\bbC)$ the mapping $A \colon \gothgl_n(\bbC) \to
\gothgl_n(\bbC)$
defined by the equality
\begin{equation}
A(x) = - h \: {}^{B\!} x \: h^{-1} \label{e:37}
\end{equation}
is an automorphism of $\gothgl_n(\bbC)$. By an inner automorphism of
$\gothgl_n(\bbC)$ generated by an element $g \in \rmGL_n(\bbC)$ the
automorphism $A$ is conjugated to an automorphism of the same form
with the replacement $h \to g \: h \: g^{-1}$, $B \to {}^{t\!}
(g^{-1}) B \: g^{-1}$. Note also that for any $g \in \rmGL_n(\bbC)$
the replacement $h \to h \: g$, $B \to  B \: g$ does not change the
automorphism $A$.

Since $h^{2N} = \nu I$ where $\nu$ is equal either to $1$ or to $-1$,
then the diagonal matrix elements of $h$ are of modulus one. In the
case when $\nu = 1$ it is convenient to represent each of them either
as $\rme^{2 \pi \rmi m / 2N}$ or as $-\rme^{2 \pi \rmi m / 2N}$, where
$m$ is an integer satisfying the condition $0 < m \le N$. In the case
when $\nu = -1$ we use either the representation $\rme^{2 \pi \rmi (m
+ 1/2) / 2N}$ or the representation $-\rme^{2 \pi \rmi (m + 1/2)/ 2N}$
where $m$ is again an integer satisfying the condition $0 < m \le N$.
Multiplying the matrices $h$ and $B$ from the right by the appropriate
diagonal matrix we exclude the second variant of the representation.
Then using inner automorphisms of $\gothgl_n(\bbC)$ we order the
diagonal elements of $h$ in a convenient way. There are two different
cases corresponding to two additional types of $\bbZ_M$-gradations of
$\gothgl_M(\bbC)$.

\subsubsection*{Type II}

Assume that there is no diagonal matrix elements of $h$ equal to $1$.
In this case one succeeds in bringing $h$ to the form (\ref{e:9})
where $\mu_\alpha = \rme^{2 \pi \rmi m_\alpha / 2N}$ if $\nu = 1$,
or $\mu_\alpha = \rme^{2 \pi \rmi (m_\alpha + 1/2) / 2N}$ if
$\nu = -1$. In both cases the integers $m_\alpha$ satisfy the
inequality $0 < m_1 < \ldots < m_p \le N$ and the relations
\[
m_{p-\alpha+1} = N - m_\alpha, \qquad \alpha = 1, \ldots, p.
\]
The automorphism $A$ generating the $\bbZ_M$-gradation under
consideration is given now by the relation (\ref{e:37}) where $B =
K_n$. Introducing integers $k_\alpha = m_\alpha - m_{\alpha + 1}$ one
easily sees that the $\bbZ_{2N}$-gradation of $\gothgl_n(\bbC)$ under
consideration is specified by the same data as a $\bbZ_N$-gradation of
type I of the corresponding orthogonal or symplectic Lie algebra.

\subsubsection*{Type III}

Let some diagonal matrix elements of $h$ be equal to $1$. In this case
$\nu = 1$, and one brings $h$ to the form (\ref{e:9}) where
$\mu_\alpha = \rme^{2 \pi \rmi m_\alpha / 2N}$ with
$m_\alpha$ satisfying the inequality $0 < m_1 < \ldots < m_p \le N$
and the relations
\[
m_{p-\alpha+2} = N - m_\alpha, \qquad \alpha = 2, \ldots, p.
\]
The automorphism $A$ generating the $\bbZ_M$-gradation under
consideration is given now by the relation (\ref{e:37})
where\footnote{Note that in this case $n-n_1$ is even with necessity.}
\[
B = \left( \begin{array}{cc}
J_{n_1} & 0 \\
0 & K_{n-n_1}
\end{array} \right).
\]
Here $n_1$ is the number of the diagonal matrix elements of $h$ equal
to $1$. Introducing integers $k_\alpha = m_\alpha - m_{\alpha + 1}$
one sees that the $\bbZ_{2N}$-gradation of $\gothgl_n(\bbC)$ under
consideration is specified by the same data as a $\bbZ_N$-gradation of
type II of the corresponding orthogonal or symplectic Lie algebra.

\vskip .5em

After all, one can show \cite{NirRaz07} that $\bbZ_{2N}$-gradations of
$\gothgl_n(\bbC)$ of types II and III can be depicted by the scheme
given in Figure \ref{f:10}.
\begin{figure}[htb]
\[
\left( \begin{array}{c|c|c|c|c}
[0]_{2N} & [k_1]_{2N} & [k_1 + k_2]_{2N} & \cdots & \displaystyle
[\sum_{\alpha=1}^{p-1} k_\alpha]_{2N} \\
{}[N]_{2N} & [k_1 + N]_{2N} & [k_1 + k_2 + N]_{2N} & \cdots &
\displaystyle [\sum_{\alpha=1}^{p-1} k_\alpha + N]_{2N} \\\hline
-[k_1]_{2N} & [0]_{2N} & [k_2]_{2N} & \cdots & \displaystyle
\sum_{\alpha=2}^{p-1}
[k_\alpha]_{2N} \\
-[k_1 + N]_{2N} & [N]_{2N} & [k_2 + N]_{2N} & \cdots & \displaystyle
[\sum_{\alpha=2}^{p-1} k_\alpha + N]_{2N} \\ \hline
-[k_1 + k_2]_{2N} & - [k_2]_{2N} & [0]_{2N} & \cdots & \displaystyle
[\sum_{\alpha = 3}^{p-1} k_\alpha]_{2N} \\
-[k_1 + k_2 + N]_{2N} & - [k_2 + N]_{2N} & [N]_{2N} & \cdots &
\displaystyle
[\sum_{\alpha = 3}^{p-1} k_\alpha + N]_{2N} \\ \hline
\vdots & \vdots & \vdots & \ddots & \vdots \\ \hline
\displaystyle - [\sum_{\alpha = 1}^{p-1} k_\alpha]_{2N} & -
\displaystyle
[\sum_{\alpha=2}^{p-1} k_\alpha]_{2N} & -\displaystyle
[\sum_{\alpha=3}^{p-1}
k_\alpha]_{2N} & \cdots & [0]_{2N} \\
{} \displaystyle -[\sum_{\alpha = 1}^{p-1} k_\alpha + N]_{2N} &
\displaystyle - [\sum_{\alpha=2}^{p-1} k_\alpha + N]_{2N} &
\displaystyle - [\sum_{\alpha=3}^{p-1}
k_\alpha + N]_{2N} & \cdots & [N]_{2N}
\end{array} \right)
\]
\caption{The structure of an outer $\bbZ_{2N}$-gradation of
$\gothgl_n(\bbC)$} \label{f:10}
\end{figure}
Here the pairs of elements of the ring $\bbZ_{2N}$ are the possible
grading indices of the corresponding blocks in the block matrix
representation (\ref{e:10}) of a general element of $\gothgl_n(\mathbb
C)$. The two possibilities are distinguished by the additional
restrictions imposed on the blocks. Namely, when the
grading index $k$ is within the range $0 \le k < N$ then
$x_{\alpha \beta} = -({}^{B\!} x)_{\alpha \beta}$
for $\alpha \le \beta$, and
$x_{\alpha \beta} = ({}^{B\!} x)_{\alpha \beta}$
for $\alpha > \beta$, while for $k$ from the range
$N \le k < 2N$ one has
$x_{\alpha \beta} = ({}^{B\!} x)_{\alpha \beta}$
when $\alpha \le \beta$, and
$x_{\alpha \beta} = -({}^{B\!} x)_{\alpha \beta}$
if $\alpha > \beta$.

\section{Explicit form of Toda equations associated with loop groups
of complex classical Lie groups}

In this section we describe the explicit form of the Toda equation
associated with loop groups of the complex classical Lie groups.
Actually we describe the explicit form of the Toda equation
(\ref{e:5}) which is equivalent to the genuine Toda equation
(\ref{e:1}) for the case of integrable $\bbZ$-gradations with finite
dimensional grading subspaces.

Let $G$ be a complex classical Lie group and $\gothg$ its Lie algebra.
Choose some $\bbZ_M$-gradation of $\gothg$. As was demonstrated above,
up to conjugations, any $\bbZ_M$-gradation of a complex classical Lie
algebra has the structure depicted either by Figure \ref{f:1} or by
Figure \ref{f:10}. In all the cases the Lie algebra $\gothg_{[0]_M}$
and the Lie group are formed by block diagonal matrices, and one can
parameterize the mapping $\gamma$ as
\begin{equation}
\psset{xunit=2.0em, yunit=1.4em}
\gamma = \left( \raise -1.6\psyunit \hbox{\begin{pspicture}(.7,.8)(4.4,4.4)
\rput(1,4){$\Gamma_1$} \rput(2,3){$\Gamma_2$} \psddots(3,2)
\rput(4,1){$\Gamma_p$}
\end{pspicture}} \right), \label{e:12}
\end{equation}
where for each $\alpha = 1, \ldots, p$ the mapping $\Gamma_\alpha$
is a mapping from $\calM$ to the Lie group
$\mathrm{GL}_{n_\alpha}(\mathbb C)$. In a general case the mappings
$\Gamma_\alpha$ are not independent. They satisfy some restrictions
imposed by the structure of the group $G$.

Let $L$ be a positive integer such that the grading subspaces
$\gothg_{+[k]_M}$ and $\gothg_{-[k]_M}$ for $0 < k < L$ are trivial.
One can see that if $x \in \gothg_{+[L]_M}$, then only the blocks
$x_{\alpha,\alpha+1}$, $\alpha = 1,\ldots,p-1$, and $x_{p1}$ in the
block matrix representation (\ref{e:10}) can be different from zero.
Thus, the mapping $c_+$ has the structure given in Figure~\ref{f:2},
\begin{figure}[ht]
\centering
\begin{minipage}[t]{.425\linewidth}
\[
\psset{xunit=2.7em, yunit=1.6em}
\left( \raise -2.4\psyunit \hbox{\begin{pspicture}(.6,.5)(5.6,5.3)
\rput(1,5){$0$} \rput(2,4.92){$C_{+1}$}
\rput(2,4){$0$} \psddots(3,4)
\psddots(3,3) \psddots(4,3)
\rput(4,2){$0$} \rput(5,1.87){$C_{+(p-1)}$}
\rput(1,.94){$C_{+0}$} \rput(5,1){$0$}
\end{pspicture}} \right)
\]
\caption{The structure of the mapping  $c_+$} \label{f:2}
\end{minipage}
\hspace{.05\linewidth}
\begin{minipage}[t]{.425\linewidth}
\[
\psset{xunit=2.7em, yunit=1.6em}
\left( \raise -2.4\psyunit \hbox{\begin{pspicture}(.5,.5)(5.5,5.3)
\rput(1,5){$0$} \rput(5,4.92){$C_{-0}$}
\rput(1,4){$C_{-1}$} \rput(2,4){$0$}
\psddots(2,3) \psddots(3,3)
\psddots(3,2) \rput(4,1.87){$0$}
\rput(4,.94){$C_{-(p-1)}$} \rput(5,1){$0$}
\end{pspicture}} \right)
\]
\caption{The structure of the mapping  $c_-$} \label{f:3}
\end{minipage}
\end{figure}
where for each $\alpha = 1, \ldots, p-1$ the mapping $C_{+\alpha}$
is a mapping from $\calM$ to the space of
$n_\alpha \times n_{\alpha+1}$
complex matrices, and $C_{+0}$ is a mapping from $\calM$ to the
space of $n_p \times n_1$ complex matrices. Here it is assumed
that if some blocks among $x_{\alpha, \alpha+1}$,
$\alpha = 1, \ldots, p-1$, and $x_{p1}$ in the general block
matrix representation (\ref{e:10}) have the grading index
different from $+[L]_M$, then the corresponding blocks in the
block representation of $c_+$ are zero matrices.

Similarly, one can see that the mapping $c_-$ has the structure
given in Figure \ref{f:3}, where for each $\alpha = 1, \ldots, p-1$
the mapping $C_{-\alpha}$
is a mapping from $\calM$ to the space of
$n_{\alpha+1} \times n_\alpha$ complex matrices, and $C_{-0}$
is a mapping from $\calM$ to the space of $n_1 \times n_p$ complex
matrices. It is assumed that if some blocks among
$x_{\alpha+1, \alpha}$, $\alpha = 1, \ldots, p-1$, and $x_{1p}$
in the general block matrix representation (\ref{e:10}) have the
grading index different from $-[L]_M$, then the corresponding
blocks in the block representation of $c_-$ are zero matrices.

The conditions (\ref{e:6}) imply
\begin{equation}
\partial_+ C_{-\alpha} = 0,
\qquad \partial_- C_{+\alpha} = 0,
\qquad \alpha = 0, \ldots, p-1.
\label{e:11}
\end{equation}
Additionally, the mappings $C_{+\alpha}$ and $C_{-\alpha}$ should
satisfy some restrictions imposed by the structure of the Lie algebra
$\gothg$.

Now assume that $G$ is the Lie group $\rmGL_n(\bbC)$ and we use a
$\bbZ_M$-gradation of $\gothgl_n(\bbC)$ of type I. It is not difficult
to show that in this case the Toda equation (\ref{e:5}) for the
mapping $\gamma$ is equivalent to the following system of equations
for the mappings
$\Gamma_\alpha$:
\begin{align}
\partial_+ \left( \Gamma_1^{-1} \: \partial_- \Gamma_1^{} \right)
&= - \Gamma_1^{-1} C_{+1}^{} \: \Gamma_2^{} \: C_{-1}^{}
+ C_{-0}^{} \Gamma_p^{-1} C_{+0}^{} \Gamma_1^{},
\notag \\*
\partial_+ \left( \Gamma_2^{-1} \: \partial_- \Gamma_2^{} \right)
&= - \Gamma_2^{-1} C_{+2}^{} \: \Gamma_3^{} \: C_{-2}^{}
+ C_{-1}^{} \Gamma_1^{-1} C_{+1}^{} \Gamma_2^{},
\notag \\*
& \quad \vdots
\label{e:13} \\*
\partial_+ \left(\Gamma_{p-1}^{-1} \: \partial_-
\Gamma_{p-1}^{}\right)
&= - \Gamma_{p-1}^{-1} C_{+(p-1)}^{} \: \Gamma_p^{} \: C_{-(p-1)}^{}
+ C_{-(p-2)}^{} \Gamma_{p-2}^{-1} C_{+(p-2)}^{} \Gamma_{p-1}^{},
\notag \\*
\partial_+ \left( \Gamma_p^{-1} \: \partial_- \Gamma_p^{} \right)
&= - \Gamma_p^{-1} C_{+p}^{} \: \Gamma_1^{} \: C_{-p}^{}
+ C_{-(p-1)}^{} \Gamma_{p-1}^{-1} C_{+(p-1)}^{} \Gamma_p^{}.
\notag
\end{align}

The Toda equations associated with loop groups of $\rmSL_n(\bbC)$
in the case of $\bbZ_M$-gradations of inner type have actually the
same form (\ref{e:13}) as the Toda equations associated with loop
groups of $\rmGL_n(\bbC)$. Here the mappings $\Gamma_\alpha$
should satisfy the condition
$\prod_{\alpha=1}^p \det \Gamma_\alpha = 1$.
If there is a solution of a Toda equation associated with a loop
group of $\rmGL_n(\bbC)$ one can easily obtain a solution of the
Toda equation associated with the corresponding loop group of
$\rmSL_n(\bbC)$. Every solution of a Toda equation associated
with a loop group of $\rmSL_n(\bbC)$ can be derived from a
solution of the Toda equation associated with the corresponding
loop group of~$\rmGL_n(\bbC)$.

Now let $G$ be the Lie group $\rmSO_n(\bbC)$ or the Lie group
$\rmSp_n(\bbC)$ and we use a $\bbZ_M$-gradation of $\gothso_n(\bbC)$
or $\gothsp_n(\bbC)$ of type I. Here we have two different cases. For
an even $p = 2s$ the Toda equation (\ref{e:5}) is equivalent to the
system
\begin{align}
\partial_+ \left( \Gamma_1^{-1} \: \partial_- \Gamma_1^{} \right)
&= -
\Gamma_1^{-1} C_{+1}^{} \: \Gamma_2^{} \: C_{-1}^{} + C_{-0}^{} \:
{}^{J\!} \Gamma_1 \: C_{+0}^{} \Gamma_1^{},
\notag \\*
\partial_+ \left( \Gamma_2^{-1} \: \partial_- \Gamma_2^{} \right)
&= -
\Gamma_2^{-1} C_{+2}^{} \: \Gamma_3^{} \: C_{-2}^{} + C_{-1}^{}
\Gamma_1^{-1} C_{+1}^{} \Gamma_2^{},
\notag \\*
& \quad \vdots  \label{e:16} \\*
\partial_+ \left( \Gamma_{s-1}^{-1} \: \partial_- \Gamma_{s-1}^{}
\right)
&= - \Gamma_{s-1}^{-1} C_{+(s-1)}^{} \: \Gamma_s^{} \: C_{-(s-1)}^{}
+ C_{-(s-2)}^{} \Gamma_{s-2}^{-1} C_{+(s-2)}^{} \Gamma_{s-1}^{},
\notag \\*
\partial_+ \left( \Gamma_s^{-1} \: \partial_- \Gamma_s^{} \right)
&= -
\Gamma_s^{-1} C_{+s}^{} {}^{J\!} (\Gamma_s^{-1}) C_{-s} +
C_{-(s-1)}^{}
\Gamma_{s-1}^{-1} C_{+(s-1)}^{} \Gamma_s^{}, \notag
\end{align}
where $C_{+0} = - {}^{J\!} C_{+0}$, $C_{-0} = - {}^{J\!} C_{-0}$,
and $C_{+s} = - {}^{J\!} C_{+s}$, $C_{-s} = - {}^{J\!} C_{-s}$ for
the the orthogonal case, whereas
$C_{+0} = {}^{J\!} C_{+0}$, $C_{-0} = {}^{J\!} C_{-0}$,
and $C_{+s} = {}^{J\!} C_{+s}$, $C_{-s} = {}^{J\!} C_{-s}$ for
the symplectic case. It should be noted that a $\bbZ_M$-gradation of
$\gothgl_n(\bbC)$ of type II for $p=2s$ also gives the equations
(\ref{e:16}). However, in this case additionally to ${}^{J\!} C_{+0} =
-C_{+0}$ and ${}^{J\!} C_{-0} = -C_{-0}$, one has ${}^{J\!} C_{+s} =
C_{+s}$ and ${}^{J\!} C_{-s} = C_{-s}$.

Returning to the case of $\bbZ_M$-gradations of $\gothso_n(\bbC)$ or
$\gothsp_n(\bbC)$ of type I, one sees that for an odd $p = 2s - 1$ the
Toda equation (\ref{e:5}) is equivalent to the system
\begin{align}
\partial_+ \left( \Gamma_1^{-1} \, \partial_- \Gamma_1^{} \right) &= -
\Gamma_1^{-1} C_{+1}^{} \, \Gamma_2^{} \, C_{-1}^{} + C_{-0}^{} \,
{}^{J\!} \Gamma_1 \, C_{+0}^{} \Gamma_1^{}, \notag \\*
\partial_+ \left( \Gamma_2^{-1} \, \partial_- \Gamma_2^{} \right) &= -
\Gamma_2^{-1} C_{+2}^{} \, \Gamma_3^{} \, C_{-2}^{} + C_{-1}^{}
\Gamma_1^{-1} C_{+1}^{} \Gamma_2^{}, \notag \\*
& \quad \vdots \label{e:15} \\*
\partial_+ \left( \Gamma_{s-1}^{-1} \, \partial_- \Gamma_{s-1}^{}
\right)
&= - \Gamma_{s-1}^{-1} C_{+(s-1)}^{} \, \Gamma_s^{} \, C_{-(s-1)}^{} +
C_{-(s-2)}^{} \Gamma_{s-2}^{-1} C_{+(s-2)}^{} \Gamma_{s-1}^{}, \notag
\\*
\partial_+ \left( \Gamma_s^{-1} \, \partial_- \Gamma_s^{} \right) &= -
{}^{B_s\!} (C_{-(s-1)}^{} \Gamma_{s-1}^{-1} C_{+(s-1)}^{} \Gamma_s^{})
+
C_{-(s-1)}^{} \Gamma_{s-1}^{-1} C_{+(s-1)}^{} \Gamma_s^{}. \notag
\end{align}
Here ${}^{B_s}\Gamma^{}_s = \Gamma^{-1}_s$ with $B_s = J$ in the
orthogonal case and $B_s = K$ in the symplectic case. One finds also
that in the orthogonal case $C_{+s} = - {}^{J\!} C_{+s}$ and $C_{-s} =
- {}^{J\!} C_{-s}$, while in the symplectic case one obtains $C_{+s} =
{}^{J\!} C_{+s}$
and $C_{-s} = {}^{J\!} C_{-s}$. Using a $\bbZ_M$-gradation of
$\gothgl_n(\bbC)$ of type II for $p=2s-1$ one also comes to the
equations (\ref{e:15}) with $B_s = K$ and the conditions $C_{+s} = -
{}^{J\!} C_{+s}$, $C_{-s} = - {}^{J\!} C_{-s}$. A $\bbZ_M$-gradation
of $\gothgl_n(\bbC)$ of type III for $p=2s - 1$ gives the equations
equivalent to the equations (\ref{e:15}) with $B_s = J$
and $C_{+s} = {}^{J\!} C_{+s}$, $C_{-s} = {}^{J\!} C_{-s}$.

Finally, if we use $\bbZ_M$-gradations of $\gothso_n(\bbC)$ or
$\gothsp_n(\bbC)$ of type III we see that the Toda equation
(\ref{e:5}) is equivalent to the system
\begin{align}
\partial_+ \left( \Gamma_1^{-1} \: \partial_- \Gamma_1^{} \right)
&= - \Gamma_1^{-1} C_{+1}^{} \: \Gamma_2^{} \: C_{-1}^{}
+ {}^{B_1\!} (\Gamma_1^{-1} C_{+1}^{} \: \Gamma_2^{} \: C_{-1}^{}),
\notag \\*
\partial_+ \left( \Gamma_2^{-1} \: \partial_- \Gamma_2^{} \right)
&= - \Gamma_2^{-1} C_{+2}^{} \: \Gamma_3^{} \: C_{-2}^{}
+ C_{-1}^{} \Gamma_1^{-1} C_{+1}^{} \Gamma_2^{},
\notag \\*
& \quad \vdots \label{e:14} \\*
\partial_+ \left( \Gamma_{s-1}^{-1} \: \partial_- \Gamma_{s-1}^{}
\right)
&= - \Gamma_{s-1}^{-1} C_{+(s-1)}^{} \: \Gamma_s^{} \: C_{-(s-1)}^{}
+ C_{-(s-2)}^{} \Gamma_{s-2}^{-1} C_{+(s-2)}^{} \Gamma_{s-1}^{},
\notag \\*
\partial_+ \left( \Gamma_s^{-1} \: \partial_- \Gamma_s^{} \right)
&= - {}^{B_s\!} (C_{-(s-1)}^{} \Gamma_{s-1}^{-1} C_{+(s-1)}^{}
\Gamma_s^{}) + C_{-(s-1)}^{} \Gamma_{s-1}^{-1} C_{+(s-1)}^{}
\Gamma_s^{}.
\notag
\end{align}
Here ${}^{B_1}\Gamma^{}_1 = \Gamma^{-1}_1$, ${}^{B_s}\Gamma^{}_s =
\Gamma^{-1}_s$ with $B_1 = J$, $B_s = J$ in the orthogonal case and
$B_1 = K$, $B_s = K$ in the symplectic case. A $\bbZ_M$-gradation of
$\gothgl_n(\bbC)$ of type III for $p=2s - 2$ gives the equations
(\ref{e:14}) with $B_1 = J$, $B_s = K$.

It is worth noting for each type of the Toda systems considered above
that if for some $\alpha$ one has $C_{+\alpha} = 0$ or $C_{-\alpha} =
0$, then the considered system of equations becomes a Toda system
associated with the corresponding finite dimensional Lie group, see,
for example, the
papers \cite{RazSavZue99,NirRaz02}. Hence, to have equations which are
really associated with loop groups one must assume that all mappings
$C_{+\alpha}$ and $C_{-\alpha}$ are nontrivial. This is possible only
when $k_\alpha = L$ for each $\alpha = 1, \ldots, p-1$ and $M = p L$.
Without any loss of generality one can assume that $L = 1$.

The equations (\ref{e:16}), (\ref{e:15}) and (\ref{e:14}) can be
obtained from the equations (\ref{e:13}) by some folding procedure
described in Appendix \ref{a:1}. That procedure also helps to
understand that there may be only the described classes of Toda
equations associated with the loop groups under consideration.

\section{Simplest case: Non-abelian sinh-Gordon and sine-Gordon
equations}
\label{s:4.2}

Defining the mappings $\Gamma_\alpha$, $C_{+\alpha}$ and $C_{-\alpha}$
for all integer values of the index $\alpha$ with the periodicity
conditions $\Gamma_{\alpha+p} = \Gamma_\alpha$,
$C_{-(\alpha+p)} = C_{-\alpha}$ and $C_{+(\alpha+p)} = C_{+\alpha}$,
one can treat the system (\ref{e:13}) as the infinite periodic system
\[
\partial_+ (\Gamma^{-1}_\alpha \partial_- \Gamma^{}_\alpha)
= - \Gamma^{-1}_\alpha C_{+\alpha} \Gamma^{}_{\alpha+1} C_{-\alpha}
+ C_{-(\alpha-1)} \Gamma^{-1}_{\alpha-1} C_{+(\alpha-1)} \:
\Gamma^{}_\alpha.
\]
In particular, when $n = r p$ and $n_\alpha = r$,
$C_{-\alpha} = C_{+\alpha} = I_r$ one comes to the
equations~\cite{Krich81,Mikh81}
\[
\partial_+ (\Gamma^{-1}_\alpha \partial_- \Gamma^{}_\alpha)
= - \Gamma^{-1}_\alpha \Gamma^{}_{\alpha+1}
+ \Gamma^{-1}_{\alpha-1} \Gamma^{}_\alpha.
\]
Putting $p = 2$ one finds the simplest set of Toda equations
\[
\partial_+ (\Gamma^{-1}_1 \partial_- \Gamma^{}_1)
= - \Gamma^{-1}_1 \Gamma^{}_{2} + \Gamma^{-1}_2 \Gamma^{}_{1},
\qquad
\partial_+ (\Gamma^{-1}_2 \partial_- \Gamma^{}_2)
= - \Gamma^{-1}_2 \Gamma^{}_{1} + \Gamma^{-1}_1 \Gamma^{}_{2},
\]
where $\Gamma_1$ and $\Gamma_2$ are $r \times r$ matrices.
These equations are invariant with respect to the
transformations $\Gamma^{}_1 \to {}^t(\Gamma_2^{-1})$,
$\Gamma^{}_2 \to {}^t(\Gamma_1^{-1})$, where the superscript
$t$ means the usual transposition. This invariance implies the
possibility of the reduction to the case when $\Gamma^{}_1 =
{}^t(\Gamma_2^{-1}) = \Gamma$. Here the equations under consideration
are reduced to the equation
\begin{equation}
\partial_+ (\Gamma^{-1} \partial_- \Gamma) = - ({}^t\Gamma
\Gamma)^{-1} + {}^t\Gamma \Gamma. \label{er:9}
\end{equation}

For the matter of possible physical applications, it is interesting to
consider real forms of Toda equations (\ref{e:5}). Let an involutive
antiholomorphic automorphism $\sigma$ of $G$ and its Lie algebra
counterpart, an involutive antilinear automorphism $\Sigma$ of
$\gothg$ consistent with the $\bbZ$-gradation, be given. The latter
means that if $x \in \gothg_{[k]_M}$, then $\Sigma(x) \in
\gothg_{[k]_M}$. Suppose also that $\Sigma(c_+) = c_+$ and
$\Sigma(c_-) = c_-$. In this case if $\gamma$ is a solution to the
Toda equation (\ref{e:5}), then $\sigma \circ \gamma$ is also a
solution to this equation,\footnote{Here we assume that $\calM$ is
$\bbR^2$.} and consistent reduction to the case when $\sigma \circ
\gamma = \gamma$ is possible.

One can construct two non-equivalent real forms of the Toda equation
(\ref{er:9}). The first one is based on the non-compact real form
given by real matrices. In this case $\sigma(\Gamma) = \Gamma^* =
\Gamma$. For $r = 1$ putting here $\Gamma = \exp(F)$ one obtains the
sinh-Gordon equation
\[
\partial_+\partial_- F = 2 \sinh F.
\]
Another one is based on the compact real form given by unitary
matrices. In this case $\sigma(\Gamma) = (\Gamma^\dagger)^{-1} =
\Gamma$ and it is convenient to write the Toda equation (\ref{er:9})
in the form
\[
\partial_+ (\Gamma^{-1} \partial_- \Gamma) = - ({}^t\Gamma
\Gamma)^\dagger + {}^t\Gamma \Gamma.
\]
For $r = 1$ putting $\Gamma = \exp(\rmi F)$ one obtains the
sine-Gordon equation
\[
\partial_+\partial_- F = 2 \sin F.
\]

\section{Conclusions}

It is observed that there are four non-equivalent classes of Toda
equations associated with loop groups of complex classical Lie groups,
when their Lie algebras are endowed with integrable
$\bbZ$-gradations with finite dimensional grading subspaces. The first
class is presented by the equations (\ref{e:13}). It is the basic
class of the loop Toda equations with no restrictions on the mappings
$\Gamma_\alpha$ and $C_{\pm \alpha}$.

In contrast, the other three classes of loop Toda equations, presented
here by the equations (\ref{e:16}), (\ref{e:15}) and (\ref{e:14}),
arise from the equations (\ref{e:13}) when one imposes some
restrictions on the mappings $\Gamma_\alpha$ and $C_{\pm \alpha}$
coming from the specific group and algebra defining conditions and
also effected by the gradation type. It is interesting, for example,
that all these three classes of loop Toda equations are revealed for
the general linear case too, under the outer type gradations, although
there was no specifying group and algebra condition at the beginning.
The classification results are supplied further with a graphic folding
procedure, see the Appendix \ref{a:1}.

\vskip2mm
This work was supported in part by the Russian Foundation for Basic
Research under grant \#07--01--00234.

\appendix

\section{Graphic representation for loop Toda equations}
\label{a:1}

Here a graphic representation for the Toda equations associated with
loop groups is described. This representation can be useful for a
simple understanding that only four non-equivalent classes
of loop Toda equations may actually arise.

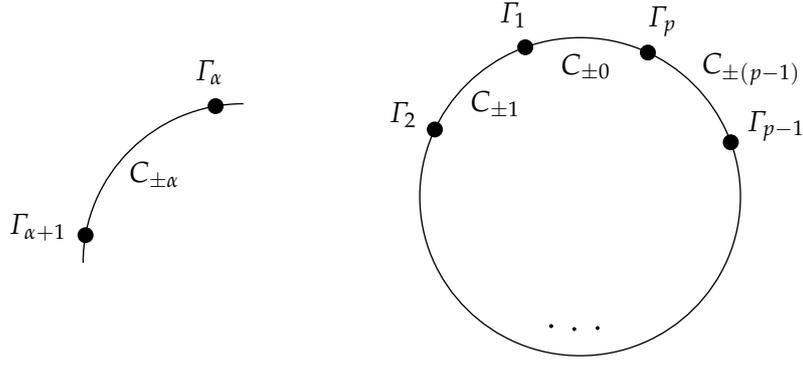
\begin{figure}[hbt]
\centering
\begin{pspicture}(-90,-50)(0,80)
\SpecialCoor
\psarc(0,0){60}{90}{180}
\psdot[dotsize=6pt](60;100)
\psdot[dotsize=6pt](60;170)
\rput(73;100){\small $\Gamma_\alpha$}
\rput(47;135){\small $C_{\pm \alpha}$}
\rput(78;170){\small $\Gamma_{\alpha+1}$}
\end{pspicture}
\qquad \qquad
\begin{pspicture}(-75,-75)(80,75)
\SpecialCoor
\pscircle(0,0){60}
\psdot[dotsize=6pt](60;20)
\psdot[dotsize=6pt](60;65)
\psdot[dotsize=6pt](60;110)
\psdot[dotsize=6pt](60;155)
\psdot[dotsize=1.5pt](50;257.5)
\psdot[dotsize=1.5pt](50;267.5)
\psdot[dotsize=1.5pt](50;277.5)
\rput(78;20){\small $\Gamma_{p-1}$}
\rput(73;65){\small $\Gamma_p$}
\rput(73;110){\small $\Gamma_1$}
\rput(73;155){\small $\Gamma_2$}
\rput(80;37){\small $C_{\pm (p-1)}$}
\rput(49;87.5){\small $C_{\pm 0}$}
\rput(47;132.5){\small $C_{\pm 1}$}
\end{pspicture}
\caption[ ]{\label{f:4} Identification rules and the general linear
case under inner type $\bbZ_M$-gradations: \\
The Toda equations (\ref{e:13}).}
\end{figure}

We use the following identification rules illustrated by
Figure \ref{f:4}.
Associate every mapping $\Gamma_\alpha$ entering the relation
(\ref{e:12}) to a small disk on a circle, and the mappings $C_{\pm
\alpha}$ to the arc between two such disks already identified with
$\Gamma_\alpha$ and $\Gamma_{\alpha+1}$. The numbering in $\alpha$ is
along the circle in the anti-clockwise direction. The whole Toda
system corresponds to a circle with $p$ small disks attached.

For the loop groups of the general linear Lie groups, when
$\bbZ_M$-gradations are of inner type, one does not have any
additional restrictions to the mappings $\gamma$, $c_+$ and $c_-$,
apart from their canonical forms given by the relation (\ref{e:12})
and Figures \ref{f:2} and \ref{f:3}. In this case, one is left with
the simple picture given in Figure \ref{f:4} depicting the Toda
equations~(\ref{e:13}).

Essentially different situations are revealed for the cases of other
classical Lie groups under inner and outer type $\bbZ_M$-gradations
and the general linear Lie group under outer
type $\bbZ_M$-gradations.

The application of the group and algebra defining conditions to the
mappings $\gamma$ and $c_+$, $c_-$, respectively, makes one fold up
the correspondingly marked circle with respect
to its diameter, the dash line in our pictures. A Toda
system is consistent with its graphic representation, if each object
on the circle identified with $\Gamma_\alpha$ and $C_{\pm \alpha}$,
$\alpha=1,\ldots,p$, finds its image counterpart on the other side
under such a folding.

\begin{figure}[hbt]
\centering
\begin{pspicture}(-75,-75)(75,75)
\SpecialCoor
\pscircle(0,0){60}
\pscircle*(60;45){3}
\pscircle*(60;135){3}
\pscircle*(60;225){3}
\pscircle*(60;270){3}
\pscircle(60;270){5}
\pscircle*(60;315){3}
\psline[linestyle=dashed, linewidth=0.3pt, dash=5pt 5pt](0,-62)(0,62)
\psline[arrows=<->](38,42.5)(-38,42.5)
\psline[arrows=<->](38,-42.5)(-38,-42.5)
\rput[lb](65;45){\small $\Gamma_{2s-1}$}
\rput[b](65;90){\small $C_{\pm 0}$}
\rput[rb](65;135){\small $\Gamma_{1}$}
\rput[rt](65;215){\small $\Gamma_{s-1}$}
\rput[rt](58;255){\small $C_{\pm (s-1)}$}
\rput[b](56;280){\small $\Gamma_s$}
\rput[lt](61;292.5){\small $C_{\pm s}$}
\rput[lt](65;325){\small $\Gamma_{s+1}$}
\psdot[dotsize=1.5pt](50;10)
\psdot[dotsize=1.5pt](50;0)
\psdot[dotsize=1.5pt](50;350)
\psdot[dotsize=1.5pt](50;170)
\psdot[dotsize=1.5pt](50;180)
\psdot[dotsize=1.5pt](50;190)
\end{pspicture}
\hspace{6em}
\begin{pspicture}(-75,-75)(75,75)
\SpecialCoor
\pscircle(0,0){60}
\pscircle*(60;45){3}
\pscircle*(60;90){3}
\pscircle(60;90){5}
\pscircle*(60;135){3}
\pscircle*(60;225){3}
\pscircle*(60;315){3}
\psline[linestyle=dashed, linewidth=0.3pt, dash=5pt 5pt](0,-62)(0,62)
\psline[arrows=<->](38,42.5)(-38,42.5)
\psline[arrows=<->](38,-42.5)(-38,-42.5)
\rput[lb](65;45){\small $\Gamma_{2s-1}$}
\rput[b](47;70){\small $C_{\pm 0}$}
\rput[rb](65;95){\small $\Gamma_{1}$}
\rput[b](47;110){\small $C_{\pm 1}$}
\rput[rb](65;135){\small $\Gamma_{2}$}
\rput[rt](65;215){\small $\Gamma_s$}
\rput[t](65;270){\small $C_{\pm s}$}
\rput[lt](65;325){\small $\Gamma_{s+1}$}
\psdot[dotsize=1.5pt](50;10)
\psdot[dotsize=1.5pt](50;0)
\psdot[dotsize=1.5pt](50;350)
\psdot[dotsize=1.5pt](50;170)
\psdot[dotsize=1.5pt](50;180)
\psdot[dotsize=1.5pt](50;190)
\end{pspicture}
\caption[ ]{\label{f:6} An odd number $p = 2s - 1$. The Toda equations
(\ref{e:15}),
where \hbox{${}^{B_s}\Gamma_s = \Gamma^{-1}_s$} and
\hbox{${}^{J}C_{\pm 0} = \varepsilon C_{\pm 0}$}, and the equivalent
Toda equations with \hbox{${}^{B_1}\Gamma_1 = \Gamma^{-1}_1$} and
\hbox{${}^{J}C_{\pm s} = \varepsilon C_{\pm s}$}. Here~$\varepsilon$
is either $+$ or $-$.}
\end{figure}
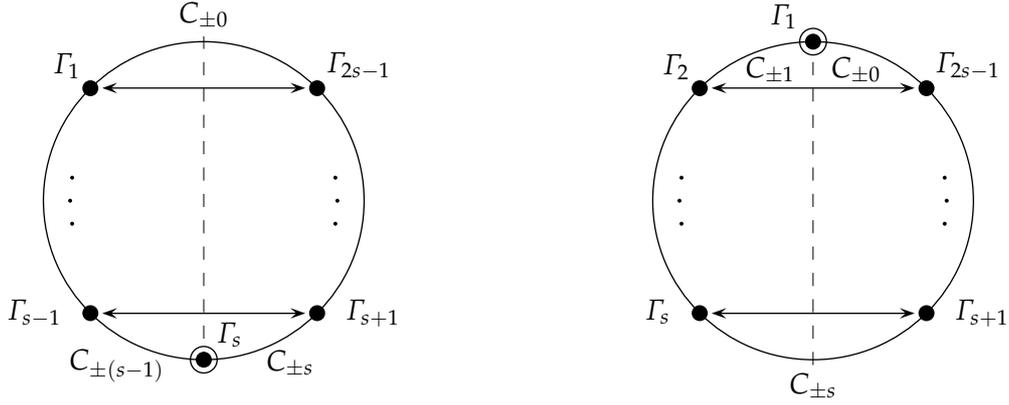

It is clear that there are only three principally different
possibilities to arrange a picture for all partitions: two for an even
$p$, Figures \ref{f:7} and \ref{f:5},
\begin{figure}[ht]
\centering
\begin{minipage}[t]{.495\linewidth}
\centering
\begin{pspicture}(-75,-75)(75,75)
\SpecialCoor
\pscircle(0,0){60}
\pscircle*(60;45){3}
\pscircle*(60;135){3}
\pscircle*(60;225){3}
\pscircle*(60;315){3}
\psline[linestyle=dashed, linewidth=0.3pt, dash=5pt 5pt](0,-62)(0,62)
\psline[arrows=<->](38,42.5)(-38,42.5)
\psline[arrows=<->](38,-42.5)(-38,-42.5)
\rput[lb](65;45){\small $\Gamma_{2s}$}
\rput[b](65;90){\small $C_{\pm 0}$}
\rput[rb](65;135){\small $\Gamma_{1}$}
\rput[rt](65;225){\small $\Gamma_{s}$}
\rput[t](65;270){\small $C_{\pm s}$}
\rput[lt](65;315){\small $\Gamma_{s+1}$}
\psdot[dotsize=1.5pt](50;10)
\psdot[dotsize=1.5pt](50;0)
\psdot[dotsize=1.5pt](50;350)
\psdot[dotsize=1.5pt](50;170)
\psdot[dotsize=1.5pt](50;180)
\psdot[dotsize=1.5pt](50;190)
\end{pspicture}
\caption[ ]{\label{f:7} An even number
$p = 2s$. The Toda equations (\ref{e:16})
where \hbox{${}^{J}C_{\pm 0} = \varepsilon C_{\pm 0}$}
and \hbox{${}^{J}C_{\pm s} = \varepsilon C_{\pm s}$}.}
\end{minipage}
\begin{minipage}[t]{.495\linewidth}
\centering
\begin{pspicture}(-75,-75)(75,75)
\SpecialCoor
\pscircle(0,0){60}
\pscircle*(60;45){3}
\pscircle*(60;90){3}
\pscircle(60;90){5}
\pscircle*(60;135){3}
\pscircle*(60;225){3}
\pscircle*(60;270){3}
\pscircle(60;270){5}
\pscircle*(60;315){3}
\psline[linestyle=dashed, linewidth=0.3pt, dash=5pt 5pt](0,-62)(0,62)
\psline[arrows=<->](38,42.5)(-38,42.5)
\psline[arrows=<->](38,-42.5)(-38,-42.5)
\rput[lb](65;45){\small $\Gamma_{2s-2}$}
\rput[b](47;70){\small $C_{\pm 0}$}
\rput[rb](65;95){\small $\Gamma_{1}$}
\rput[b](47;110){\small $C_{\pm 1}$}
\rput[rb](65;135){\small $\Gamma_{2}$}
\rput[rt](65;215){\small $\Gamma_{s-1}$}
\rput[rt](58;255){\small $C_{\pm (s-1)}$}
\rput[b](56;280){\small $\Gamma_s$}
\rput[lt](61;292.5){\small $C_{\pm s}$}
\rput[lt](65;325){\small $\Gamma_{s+1}$}
\psdot[dotsize=1.5pt](50;10)
\psdot[dotsize=1.5pt](50;0)
\psdot[dotsize=1.5pt](50;350)
\psdot[dotsize=1.5pt](50;170)
\psdot[dotsize=1.5pt](50;180)
\psdot[dotsize=1.5pt](50;190)
\end{pspicture}
\caption[ ]{\label{f:5} An even number
$p = 2s - 2$. The Toda equations (\ref{e:14}.)
where \hbox{${}^{B_1}\Gamma_1 = \Gamma^{-1}_1$}
and \hbox{${}^{B_s}\Gamma_s = \Gamma^{-1}_s$}}
\end{minipage}
\end{figure}
and one for an odd $p$, the left picture in Figure \ref{f:6}.
The additionally encircled small disks are ones which are folded up
with themselves. The right picture in Figure \ref{f:6} describes
another variant of folding. It corresponds to the Toda equations which
can be obtained from the equations (\ref{e:15}) by the substitution
$\Gamma_\alpha \to {}^{B_s\!}(\Gamma_{s-\alpha+1}^{-1})$, $C_{\pm
\alpha} \to {}^{J\!} C_{\pm(s - \alpha)}$ supplemented by the change
$B_s \to B_1$.

Finally, it is interesting to make a comparison with the Toda systems
associated with finite dimensional Lie groups
\cite{RazSavZue99,NirRaz02}. Considering a similar graphic
representation on a line for such Toda systems and taking into account
that no periodicity is at hand, one can see that only two different
classes of Toda systems, one for an even and one for an odd number
$p$, associated with the orthogonal and symplectic Lie groups can
arise there in addition to the general linear case.


\begin{thebibliography}{**}

\bibitem{LezSav92}
A. N. Leznov and M. V. Saveliev,
{\em Group-theoretical Methods for Integration of Nonlinear Dynamical
Systems\/}
(Birkh\"auser, Basel, 1992).

\bibitem{RazSav94}
A.~V.~Razumov~and~M.~V.~Saveliev,
{\em Differential geometry of Toda systems},
Commun. Abal. Geom. {\bf 2} (1994) 461--511
{\tt [arXiv:hep-th/9612081]}.

\bibitem{RazSav97a}
A.~V.~Razumov~and~M.~V.~Saveliev,
{\em Lie Algebras, Geometry, and Toda-type Systems\/}
(Cambridge University Press, Cambridge, 1997).

\bibitem{Lez85}
A.~N.~Leznov,
The internal symmetry group and methods of field theory for
integrating exactly soluble dynamic systems,
In: {\em Group Theoretical Methods in Physics\/}, Proc. of the 1982
Zvenigorod seminar (New York, Harwood, 1985), 443--457.

\bibitem{GerSav95}
J.--L.~Gervais and M.~V.~Saveliev,
{\em Higher grading generalizations of the Toda systems\/},
Nucl. Phys. B {\bf 453} (1995) 449--476
{\tt [arXiv:hep-th/9505047]}.

\bibitem{FerGerSanSav96}
L.~A.~Ferreira, J.~L.~Gervais, J.~S\'anchez Guill\'en and
M.~V.~Saveliev,
{\em Affine Toda systems coupled to matter fields\/},
Nucl. Phys. B {\bf470} (1996) 236--290
{\tt [arXiv:hep-th/9512105]}.

\bibitem{BueFerRaz02}
A. G. Bueno, L. A. Ferreira and A. V. Razumov,
{\em Confinement and soliton solutions in the SL$(3)$ Toda model
coupled to matter fields\/},
Nucl. Phys. B {\bf 626} (2002) 463--499
{\tt [arXiv:hep-th/0105078]}.

\bibitem{RazSav97c}
A.~V.~Razumov and M.~V.~Saveliev,
On some class of multidimensional nonlinear integrable systems,
In: {\it Second International Sakharov Conference in Physics\/}, eds.
I.~M.~Dremin and A.~M.~Semikhatov (World Scientific, Singapore, 1997)
547--551
{\tt [arXiv:hep-th/9607017]}.

\bibitem{RazSav97d}
A.~V.~Razumov and M.~V.~Saveliev,
{\em Multi-dimensional Toda-type systems\/},
Theor. Math. Phys. {\bf 112} (1997) 999--1022
{\tt [arXiv:hep-th/9609031]}.

\bibitem{FerMirGui95}
L. A. Ferreira, J. L. Miramontes and J. S. Guill\'en,
{\em Solitons, $\tau$-functions and hamiltonian reduction
for non-Abelian conformal affine Toda theories\/},
Nucl. Phys. B {\bf 449} (1995) 631--679
{\tt [arXiv:hep-th/9412127]}.

\bibitem{FerGalHolMir97}
C. R. Fern\'andez-Pousa, M. V. Gallas, T. J. Hollowood
and J. L. Miramontes,
{\em The symmetric space and homogeneous sine-Gordon theories\/},
Nucl. Phys. B {\bf 484} (1997) 609--630
{\tt [arXiv:hep-th/9606032]}.

\bibitem{RazSav97b}
A.~V.~Razumov and M.~V.~Saveliev,
{\em Maximally non-abelian Toda systems},
Nucl. Phys. B {\bf 494} (1997) 657--686
{\tt [arXiv:hep-th/9612081]}.

\bibitem{RazSavZue99}
A.~V.~Razumov, M.~V.~Saveliev and A.~B.~Zuevsky,
Non-abelian Toda equations associated with classical Lie groups,
In: {\em Symmetries and Integrable Systems\/}, Proc. of the
Seminar, ed. A. N. Sissakian (JINR, Dubna, 1999) 190--203
{\tt [arXiv:math-ph/9909008]}.

\bibitem{NirRaz02}
Kh.~S.~Nirov~and~A.~V.~Razumov,
On classification of non-abelian Toda systems,
In: {\em Geometrical an Topological Ideas in Modern Physics\/},
ed. V. A. Petrov (IHEP, Protvino, 2002) 213--221
{\tt [arXiv:nlin.SI/0305023]}.

\bibitem{PreSea86}
A.~Pressley~and~G.~Segal,
{\em Loop Groups\/}
(Clarendon Press, Oxford, 1986).

\bibitem{NirRaz06}
Kh.~S.~Nirov~and~A.~V.~Razumov,
{\em On $\bbZ$-gradations of twisted loop Lie algebras of complex
simple Lie algebras\/},
Commun. Math. Phys. {\bf 267} (2006) 587--610
{\tt [arXiv:math-ph/0504038]}.

\bibitem{NirRaz07}
Kh.~S.~Nirov~and~A.~V.~Razumov,
{\em Toda equations associated with loop groups of complex classical
Lie groups\/}, Nucl. Phys. {\bf B}, to appear (2007)
{\tt [arXiv:math-ph/0612054]}.

\bibitem{Ham82}
R.~Hamilton,
{\em The inverse function theorem of Nash and Moser\/},
Bull. Am. Math. Soc. {\bf 7} (1982) 65--222.

\bibitem{Mil84}
J.~Milnor,
{\em Remarks on infinite-dimensional Lie groups\/},
In: Relativity, Groups and Topology II, eds. B. S. DeWitt and
R. Stora (North-Holland, Amsterdam, 1984) p. 1007--1057.

\bibitem{OniVin90}
A.~L.~Onishchik~and~E.~B.~Vinberg,
{\em Lie Groups and Algebraic Groups\/},
(Springer, Berlin, 1990).

\bibitem{Kac94}
V.~G.~Kac,
{\em Infinite dimensional Lie algebras\/}, 3rd ed.
(Cambridge University Press, Cambridge, 1994).

\bibitem{GorOniVin94}
V.~V.~Gorbatsevich,~A.~L.~Onishchik~and~E.~B.~Vinberg,
Structure of Lie Groups and Lie Algebras, In:
{\em Lie Groups and Lie Algebras III\/},
Encyclopaedia of Mathematical Sciences, v. 41, (Springer, Berlin,
1994).

\bibitem{Krich81}
I.~M.~Krichever,
{\em The periodic non-abelian Toda chain and its two-dimensional
generalization\/},
Russ. Math. Surv. {\bf 36} (1981) 82--89.

\bibitem{Mikh81}
A.~V.~Mikhailov,
{\em The reduction problem and the inverse scattering method\/},
Physica {\bf 3D} (1981) 73--117.


\end{thebibliography}
\end{document}